\documentclass[12pt]{article}
\pdfoutput=1
\usepackage{graphicx,psfrag,epsf,color}
\usepackage{amsmath,amssymb,amsfonts}
\usepackage{array}
\usepackage{cite}
\usepackage{rotating}
\usepackage{slashed, cancel}
\usepackage{xcolor}
\usepackage[caption=false]{subfig}
\usepackage{graphicx}
\usepackage{mathrsfs}  
\usepackage{tabularx}
\usepackage{multirow} \usepackage{array}
\usepackage{floatrow}

\usepackage{color,xcolor}
\usepackage{lscape}
\usepackage[colorlinks,citecolor=blue,urlcolor=blue,linkcolor=blue]{hyperref}

\newcolumntype{C}{>{\centering\arraybackslash}X}
\numberwithin{equation}{section}
\newcommand{\alem}{\alpha_{\rm em}}
\newcommand{\reg}{\alpha}
\newcommand{\eps}{\epsilon}
\newcommand{\bra}[1]{\big\langle{#1}\big\vert}
\newcommand{\ket}[1]{\big\vert{#1}\big\rangle}

\newcommand{\scetone}{SCET-1}
\newcommand{\scettwo}{SCET-2}
\DeclareSymbolFont{matha}{OML}{txmi}{m}{it}
\DeclareMathSymbol{\varv}{\mathord}{matha}{118}
\DeclareFloatFont{small}{\footnotesize}

\begin{document}
\allowdisplaybreaks

\begin{titlepage}

\begin{flushright}
{\small
MITP-22-034\\
SI-HEP-2022-10\\
P3H-22-049\\
May 13, 2022
}
\end{flushright}

\vskip1cm
\begin{center}
{\Large \bf\boldmath Muon-electron backward scattering:\\ [0.4em]
a prime example for endpoint singularities in SCET}
\end{center}

\vspace{0.5cm}
\begin{center}
Guido~Bell$^a$, Philipp B\"oer$^b$ 
and Thorsten Feldmann$^a$ \\[6mm]

{\it $^a$Theoretische Physik 1, Center for Particle Physics Siegen, \\
Universit\"at Siegen, 57068 Siegen, Germany}\\[0.3cm]

{\it $^b$PRISMA$^+$ Cluster of Excellence \& Mainz Institute for Theoretical Physics,\\ 
Johannes Gutenberg Universit\"at, 55099 Mainz, Germany}
\end{center}

\vspace{0.6cm}
\begin{abstract}
\vskip0.2cm\noindent

We argue that energetic muon-electron scattering in the backward direction can be viewed as a template case to study the resummation of large logarithms related to endpoint divergences appearing in the effective-theory formulation of hard-exclusive processes. While it is known since the mid sixties that the leading double logarithms from QED corrections resum to a modified Bessel function on the amplitude level, the modern formulation in Soft-Collinear Effective Theory (SCET)  shows a surprisingly complicated and iterative pattern of endpoint-divergent convolution integrals. In contrast to the bottom-quark induced $h \to \gamma\gamma$ decay, for which a renormalized factorization theorem has been proposed recently, we find that rapidity logarithms generate an infinite tower of collinear-anomaly exponents. This can be understood as a generic consequence of the underlying $2\to 2$ kinematics.
Using endpoint refactorization conditions for the collinear matrix elements, we show how the Bessel function is reproduced in the effective theory from consistency relations between quantities in a ``bare'' factorization theorem.
\end{abstract}

\end{titlepage}

\section{Introduction}
\label{sec:Intro}

In the past two decades Soft-Collinear Effective Theory (SCET)~\cite{Bauer:2000yr,Bauer:2001yt,Beneke:2002ph,Beneke:2002ni} has been applied in different branches of particle physics to derive factorization theorems and to resum logarithmically-enhanced corrections to all orders in perturbation theory. Somewhat surprisingly, by far the largest fraction of the SCET analyses has focused  exclusively on the leading terms in the underlying power expansion. With the ever increasing precision of perturbative QCD calculations, the study of next-to-leading power corrections has started only recently to attract considerable attention (see e.g.~\cite{Larkoski:2014bxa,Moult:2017rpl,Feige:2017zci,Chang:2017atu,Beneke:2017ztn,Moult:2018jjd,Beneke:2018rbh,Beneke:2018gvs,Ebert:2018gsn,Beneke:2019kgv,Beneke:2019mua,Moult:2019uhz,Liu:2019oav,Wang:2019mym,Beneke:2020ibj,Liu:2020tzd,Liu:2020wbn,Beneke:2022obx}). While the subleading terms in the effective Lagrangian are known since the early applications of SCET in the context of charmless $B$-meson decays~\cite{Chay:2002vy,Manohar:2002fd,Beneke:2002ph,Pirjol:2002km,Beneke:2002ni}, the recent work has focused e.g.~on constructing subleading operator bases~\cite{Moult:2017rpl,Feige:2017zci,Chang:2017atu} and on studying the anomalous dimensions of power-suppressed SCET operators~\cite{Beneke:2017ztn,Beneke:2018rbh}. Nevertheless, there currently exists only a very limited number of processes, for which the resummation of large logarithmic corrections has been extended to subleading power, which includes e.g.~the study of threshold logarithms in Drell-Yan and Higgs production~\cite{Beneke:2018gvs,Beneke:2019mua,Beneke:2020ibj}, $e^+e^-$ event-shape distributions~\cite{Moult:2018jjd,Beneke:2022obx} and the bottom-quark induced $h \to\gamma\gamma$ decay~\cite{Liu:2019oav,Liu:2020tzd,Liu:2020wbn,Wang:2019mym}.

The study of power corrections is a very non-trivial task, and among the many complications that one faces, the most challenging one seems to be the problem of endpoint-divergent convolution integrals. Due to the presence of energetic particles in the effective theory, SCET is a non-local theory, and the factorization theorems take the form of a convolution of a hard-coefficient function with matrix elements of non-local SCET operators.\footnote{Throughout this paper we use the notion ``convolution'' in a loose sense for every integration in a factorization theorem. The integrations are not necessarily convolutions in the strict mathematical sense.} As long as these convolutions are well-behaved at the endpoints, the operator matrix elements can be renormalized by means of distributions prior to taking the convolutions, and the logarithmic corrections can be resummed using renormalization-group techniques. This is, however, no longer true if the convolutions are by themselves divergent at the endpoint and produce poles in the applied regulators, which spoils the renormalization program.

The problem of endpoint-divergent convolution integrals seems to arise generically in SCET at subleading power. Further progress in this field therefore requires to better under\-stand the endpoint dynamics, and to formulate factorization theorems that hold for the renormalized rather than the bare quantities. In a series of papers~\cite{Liu:2019oav,Liu:2020tzd,Liu:2020wbn}, this has recently been achieved for the bottom-quark induced $h \to (b\bar b)^*\to\gamma\gamma$ decay amplitude. In that work the derivation of a renormalized factorization theorem proceeds via two key steps:
\begin{itemize}
\item
The presence of endpoint divergences in the bare factorization theorem signals a sensitivity to configurations that invalidate the generic scaling of the soft and collinear momenta. In the endpoint region, the bare functions therefore become multi-scale objects that must be refactorized. While this idea of \emph{endpoint refactorization} has by now been used in different contexts~\cite{Liu:2019oav,Beneke:2020ibj,Beneke:2022obx}, it was -- to the best of our knowledge -- for the first time proposed in a thesis by one of us in~\cite{Boer:2018mgl}.
\item
In a second step the terms in the bare factorization theorem are reorganized in a way that the endpoint contributions are properly subtracted in the convolution integrals. In the specific implementation of~\cite{Liu:2019oav,Liu:2020tzd,Liu:2020wbn}, this \emph{rearrangement} naturally induces cutoffs that regularize the endpoint-divergent convolutions. This step depends, however, on the specific structure of the bare factorization theorem, and it remains to be investigated if it can be applied to other processes as well. 
\end{itemize}

In the present paper we argue that the simple textbook process of muon-electron scattering in the backward direction can be viewed as a template case to study the problem of endpoint divergences in SCET. More specifically, we consider the $\mu^- e^- \to \mu^- e^-$ scattering amplitude in the exact backward limit at asymptotically large center-of-mass energies $\sqrt{s} \gg m_{e,\mu}$. It is known for more than 50 years that the dominant double-logarithmic corrections $\sim\alem^{n+1} \ln^{2n} m_{e,\mu}/\sqrt{s}$ resum to a modified Bessel function in this limit~\cite{Gorshkov:1966qd,Berestetskii:1982qgu}. The goal of the present article consists in reproducing this result with modern effective-field-theory methods, and in illustrating that this process shows a more complicated pattern of endpoint divergences than the $h \to (b\bar b)^*\to\gamma\gamma$ decay. We therefore advocate this simple QED process as a prime example for studying non-trivial aspects of soft-collinear factorization for generic $2\to2$ processes, or processes with even higher particle multiplicities. In particular, we believe that our findings can shed new light onto the factorization of hard-exclusive processes in QCD, like $B$-meson decays into energetic light hadrons. Our analysis in fact descends from a specific study of $B_c \to \eta_c$ form factors in~\cite{Boer:2018mgl,Bell:2005gw}.

The analysis of the muon-electron backward-scattering process is formulated in an effective theory that is referred to as \scettwo. Due to the presence of massive fermions in the loop integrals, the effective-theory formulation requires a rapidity regulator in addition to the usual dimensional regulator. Interestingly, it turns out that the $\mu^- e^- \to \mu^- e^-$ scattering amplitude contributes at \emph{leading power}, but the convolution integrals are nevertheless endpoint-divergent because of a specific \emph{soft-enhancement mechanism} that we describe in Sec.~\ref{sec:SCET}. Due to this mechanism certain helicity flips can occur an arbitrary number of times without inducing any type of power suppression, which is at the origin of an iterative pattern of endpoint divergences.

While we do not aim at deriving a renormalized factorization theorem in this work, we will show that the dominant double logarithms can be resummed in the effective theory, using endpoint refactorization conditions and consistency relations. Interestingly, the Bessel function is reproduced in our analysis via an infinite tower of collinear-anomaly exponents -- a structure that has not yet been observed in any previous application of \scettwo~before. The considered muon-electron scattering process therefore provides new insights into the effective theory, even on the simplest double-logarithmic level, and it allows one to clearly isolate the problem of endpoint divergences on the level of a leading-power QED process.

This article is organized as follows. In Sec.~\ref{sec:setup} we present some basic definitions, and we show how the dominant double logarithms can be resummed with ``traditional'' diagrammatic methods. In Sec.~\ref{sec:SCET} we derive a bare factorization theorem for the backward-scattering amplitude, and we illustrate that it suffers from endpoint-divergent convolution integrals. We then derive endpoint refactorization conditions for the collinear matrix elements in Sec.~\ref{sec:resummation}, and we show how the perturbative expansion of the modified Bessel function is recovered order-by-order in the effective theory from consistency relations. We finally perform a detailed comparison of the SCET analysis for the bottom-quark induced $h \to\gamma\gamma$ decay and the $\mu^- e^- \to \mu^- e^-$ backward-scattering process in Sec.~\ref{sec:hgg}, before we conclude in Sec.~\ref{sec:conclusion}. Some technical aspects of our analysis are discussed in the Appendix.

\section{Preliminaries}
\label{sec:setup}
In this section we introduce our notation to describe the kinematics in the muon-electron backward scattering. 
We identify two Dirac structures that contribute at leading power in the small expansion parameter and define the corresponding form factors. To set the stage for the effective-theory analysis, we follow the traditional diagrammatic approach to sum the leading double-logarithmic contributions, which arise from particular ladder-type diagrams with soft lepton propagators, to all orders.

\subsection{Kinematics and power counting}
\label{subsec:kinematics}

We consider the $2\to2$ scattering process \begin{align} 
 e^-(p_1) \, \mu^-(p_2) \longrightarrow e^-(p'_1) \, \mu^-(p'_2)
\end{align}
of massive leptons, $p_1^2 = {p'_1}^2 = m_e^2$ and $p_2^2 = {p'_2}^2 = m_\mu^2$, at large center-of-mass energies $\sqrt{s} \gg m_{e,\mu}$. In the kinematic situation of exact backward scattering all four external leptons can be chosen to move along the $z$-axis. We then define the standard light-cone vectors $n_\pm^\mu$, with $n_+^2 = n_-^2 = 0$ and $n_+ n_- = 2$, such that the external momentum components perpendicular to $n_\pm^\mu$ vanish, and we can write
\begin{align}
    p^{(\prime)}_i{}^\mu = \frac{n_+ p^{(\prime)}_i}{2} n_-^\mu + \frac{n_- p^{(\prime)}_i}{2} n_+^\mu \,.
\end{align}
The longitudinal components can be expressed as
\begin{align}
 n_+p_1 &= n_-p'_1 =  \frac{s+m_e^2-m_\mu^2+\sqrt{\lambda(s,m_e^2,m_\mu^2)}}{2\sqrt{s}} = \frac{\sqrt{s}-\sqrt{u}+\sqrt{-t}}{2} \,, \nonumber \\
 n_-p_2 &= n_+p'_2 =  \frac{s-m_e^2+m_\mu^2+\sqrt{\lambda(s,m_e^2,m_\mu^2)}}{2\sqrt{s}} = \frac{\sqrt{s}+\sqrt{u}+\sqrt{-t}}{2} \,,
\end{align}
with the Mandelstam variables $s,t,u$ and the K\"all\'en function $\lambda(s,m_e^2,m_\mu^2) = s^2+m_e^4+m_\mu^4- 2m_e^2 s -2 m_\mu^2 s - 2m_e^2 m_\mu^2$. The remaining light-cone components are  then fixed by the on-shell conditions \mbox{$(n_-p^{(\prime)}_1)(n_+p^{(\prime)}_1) = m_e^2$} and \mbox{$(n_-p^{(\prime)}_2)(n_+p^{(\prime)}_2) = m_\mu^2$}. In the exact backward limit, $s$ is the only independent Mandelstam variable, whereas \mbox{$t=-\lambda(s,m_e^2,m_\mu^2)/s$} and $u=(m_\mu^2-m_e^2)^2/s$ are fixed by $s$ and the lepton masses.

We are interested in the asymptotic limit of the QED scattering amplitude at large energies, $s\gg m^2_{\mu} \gtrsim m^2_{e}$. In this limit the amplitude can be simplified by an expansion in the small power-counting parameter $\lambda = m_{\mu}/\sqrt{s}$, and we aim at resumming double-logarithmic corrections $\sim\alem^{n+1} \ln^{2n} m_{e,\mu}/\sqrt{s}$ to all orders in perturbation theory using the framework of SCET.\footnote{The expansion parameter $\lambda$ is not to be confused with the K\"all\'en function $\lambda(s,m_e^2,m_\mu^2)$, which we always indicate by the additional kinematic arguments.} We emphasize that, although the mass ratio $m_e/m_\mu$ is numerically small, we count $m_e/m_\mu=\mathcal{O}(1)$ since the ratio does not scale with the center-of-mass energy $\sqrt{s}$. This implies that we do not resum any logarithms of the form $\alem \ln^2 m_e/m_\mu$. With respect to the power-counting parameter $\lambda$, the external momenta then scale in their light-cone components $(n_+p^{(\prime)}_i,(p^{(\prime)}_i)_\perp,n_-p^{(\prime)}_i)$ as
\begin{align}\label{eq:externalmom}
    p_1^\mu \sim p'_2{}^\mu \sim (1,0,\lambda^2)\sqrt{s} \,, \qquad 
		p_2^\mu \sim p'_1{}^\mu \sim (\lambda^2,0,1)\sqrt{s} \,,
\end{align}
which, as we will explain in more detail in Sec.~\ref{sec:SCET}, amounts to the respective scaling of collinear and anti-collinear modes in SCET. We remark that the kinematic configuration implies a certain fine-tuning of the momenta, in the sense that the difference of the two (anti-)collinear momenta must be ultra-soft,
\begin{align}
    p_2^\mu - p'_1{}^\mu = p'_2{}^\mu - p_1^\mu = \sqrt{u} \; \frac{n_-^\mu+n_+^\mu}{2} \sim (\lambda^2,0,\lambda^2)\sqrt{s} \,,
\end{align}
with invariant mass $(p_2 - p'_1)^2 = (p'_2 - p_1)^2 = u =\mathcal{O}(\lambda^4 s) \ll m_{e,\mu}^2$.

Due to the counting $m_e/m_\mu=\mathcal{O}(1)$, it is sufficient to consider the equal-mass limit $m_e = m_\mu \equiv m$ for the resummation of 
the leading double-logarithmic corrections.\footnote{We consider distinguishable lepton flavours to avoid cross diagrams from the exchange of identical particles.} The kinematics then simplifies to 
\begin{align}
 n_+p_1 = n_-p'_1 = n_-p_2 = n_+p'_2 =  \frac{\sqrt{s}}{2} \left(1+\sqrt{1-\frac{4m^2}{s}}\right) = \sqrt{s} \;\Big(1+ \mathcal{O}(\lambda^2)\Big) \,,
\end{align}
which implies that $p_1^\mu = p'_2{}^\mu \equiv p^\mu$ and $p_2^\mu = p'_1{}^\mu \equiv \bar{p}^\mu$, and the problem becomes symmetric with respect to the exchange of the light-cone directions $n_-^\mu \leftrightarrow n_+^\mu$. The Mandelstam variables furthermore simplify to $t = 4m^2-s \approx -s$ and $u = 0$ in this approximation. Obviously, once one is interested in resumming subleading logarithms, one cannot use these simplifications any longer. Nevertheless, as we will see below, already the double logarithms of the leading-power amplitude have an interesting all-order structure that shares many complications related to endpoint singularities in SCET.
In order to keep the discussion transparent, we therefore adopt the equal-mass limit in the following.

\subsection{Fixed-order analysis}
\label{subsec:fixed-order}

\begin{figure}
\centering
\includegraphics[scale=2.2]{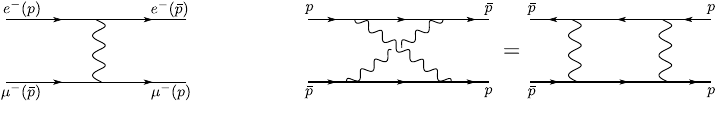}
	\caption{Left: Tree-level exchange of a hard photon
	between the collinear and anti-collinear leptons. 
	Right: One-loop graph that contributes to the amplitude at the double-logarithmic level (in Feynman gauge). We find it convenient to display this diagram with a twisted electron line, such that all (anti-)collinear external momenta are on the (left) right side of the diagram.}
	\label{fig:treeandnlo}
\end{figure}

At lowest order in perturbation theory, the backward scattering happens through the $t$-channel exchange of a hard photon, as depicted in the left panel of Fig.~\ref{fig:treeandnlo}. The corresponding amplitude reads
\begin{align}
 {\cal M}^{(0)} &= \frac{4\pi\alem}{t} \, \big[\bar{u}^{(e)}(\bar p)\gamma^\nu u^{(e)}(p)\big] \, \big[\bar{u}^{(\mu)}(p) \gamma_\nu u^{(\mu)}(\bar p)\big] \,.
\end{align}
In each Dirac string the matrix $\gamma^\nu$ is sandwiched between a collinear and an anti-collinear spinor. At leading power in $\lambda$, the equations of motion imply that only the perpendicular component
\mbox{$\gamma_\perp^\mu = \gamma^\mu - \frac{\slashed{n}_-}{2} n_+^\mu - \frac{\slashed{n}_+}{2} n_-^\mu$}
contributes,
\begin{align}
\label{eq:LOlp}
 {\cal M}^{(0)} &= \frac{4\pi\alem}{-s} \, \big[\bar{u}^{(e)}_{\bar{\xi}}\gamma_\perp^\nu u_\xi^{(e)}\big] \, \big[\bar{u}_\xi^{(\mu)} \gamma_{\perp\nu} u^{(\mu)}_{\bar{\xi}}\big] + \mathcal{O}(\lambda) \,,
\end{align}
and the tree-level amplitude can be expressed in terms of a single Dirac structure. Here $u_\xi$ and $u_{\bar{\xi}}$ denote the large components of the respective on-shell spinors of collinear and anti-collinear particles in the large-energy limit, which fulfill the equations of motion \mbox{$\slashed{n}_- u_{\xi} = 0 = \slashed{n}_+ u_{\bar{\xi}}$}. Performing a Fierz transformation, we can write
\begin{align}
\label{eq:Fierz}
    \big[\bar{u}^{(e)}_{\bar{\xi}} \gamma_\perp^\nu u^{(e)}_\xi\big] \, \big[\bar{u}_\xi^{(\mu)} \gamma_{\perp\nu} u_{\bar{\xi}}^{(\mu)}\big] &= -2 \, \big[\bar{u}^{(e)}_{\bar{\xi}} \frac{\slashed{n}_-}{2} P_R \, u^{(\mu)}_{\bar{\xi}}\big] \, \big[\bar{u}_\xi^{(\mu)} \frac{\slashed{n}_+}{2} P_R \, u^{(e)}_\xi\big] + (R \leftrightarrow L)\,,
\end{align}
where $P_{L,R}=\frac12(1\mp\gamma_5)$ are the usual chiral projection operators. This expression shows that the chiralities of the leptons are conserved and decouple,
\begin{align}
    {\cal M}^{(0)}(e^-\mu^- \to e^-\mu^-) = {\cal M}^{(0)}(e_R^-\mu_R^- \to e_R^-\mu_R^-) + {\cal M}^{(0)}(e_L^-\mu_L^- \to e_L^-\mu_L^-) + \mathcal{O}(\lambda) \,,
\end{align}
as expected in the high-energy limit.

Radiative corrections generate a form factor that multiplies the tree-level amplitude ${\cal M}^{(0)}$, but they also induce another helicity-flipping Dirac structure
\begin{align}
    \widetilde{\cal M} &= \frac{4\pi\alem}{-s} \left(\big[\bar{u}^{(e)}_{\bar{\xi}} u^{(e)}_\xi\big] \, \big[\bar{u}^{(\mu)}_\xi u^{(\mu)}_{\bar{\xi}}\big]-\big[\bar{u}^{(e)}_{\bar{\xi}}\gamma_5 u^{(e)}_\xi\big] \,\big[\bar{u}^{(\mu)}_\xi \gamma_5 u^{(\mu)}_{\bar{\xi}}\big]\right) \nonumber \\
    &= \frac{4\pi\alem}{-s} \Big(2 \,  \big[\bar{u}^{(e)}_{\bar{\xi}} \frac{\slashed{n}_-}{2} P_R \, u^{(\mu)}_{\bar{\xi}}\big] \, \big[\bar{u}_\xi^{(\mu)} \frac{\slashed{n}_+}{2} P_L \, u^{(e)}_\xi\big] + (R \leftrightarrow L) \Big) \,,
\end{align}
at leading power in $\lambda$. In general, the amplitude can thus be described in terms of two form factors in the high-energy limit, a helicity-conserving form factor $F_1(\lambda)$ and a helicity-changing form factor $F_2(\lambda)$,
\begin{align}
    {\cal M} = F_1(\lambda) \,{\cal M}^{(0)} + F_2(\lambda) \,\widetilde{{\cal M}} \,+\, \mathcal{O}(\lambda) \,.
\end{align}
In the following we expand these form factors perturbatively as $F_i(\lambda) = \sum_{n=0}^\infty \left(\frac{\alem}{2\pi}\right)^n F_i^{(n)}(\lambda)$ with $F_1^{(0)}(\lambda)=1$ and $F_2^{(0)}(\lambda)=0$ for the tree-level contributions.

At one-loop order one needs to evaluate the usual box, vertex and self-energy diagrams for a four-fermion scattering amplitude. We find that the next-to-leading order (NLO) corrections to the form factors are given at leading power in $\lambda$ by
\begin{align}
\label{eq:F1oneloop}
    F_1^{(1)}(\lambda) &= 
		\frac12 \ln^2\lambda^2 -\ln\lambda^2 - \frac{2 n_\ell}{3} \ln\frac{\mu^2}{s}
		- \frac{10 n_\ell}{9} - \frac{\pi^2}{3} -4
		+ 2\pi i \left(\frac{1}{\eps_{\rm IR}} + \ln \frac{\mu^2}{s}\right) \,,
\end{align}
with $n_\ell$ the number of lepton flavours, and
\begin{align}
    F_2^{(1)}(\lambda) &= \ln \lambda^2 + 2 \,,
\end{align}
where we renormalized the QED coupling in the $\overline{\text{MS}}$-scheme. 

Several comments about this result are in order. First, we observe that the helicity-changing contributions in $F_2(\lambda)$ are at most single-logarithmic, and they in fact do not contribute in the interference with the Born amplitude on the cross-section level. The double logarithms $\sim\alem^{n} \ln^{2n} m/\sqrt{s}$, which are the concern of our paper, are thus entirely captured by the helicity-conserving form factor $F_1(\lambda)$, on which we will concentrate in the following. Second, the amplitude contains a soft singularity, which again is of lower logarithmic order and therefore irrelevant for the following discussion. As usual, the $1/\eps_{\rm IR}$-divergence in the amplitude is cancelled by the corresponding real-emission process in a soft-photon inclusive observable. The last and for our purpose most important comment about the result in~\eqref{eq:F1oneloop} is that -- in Feynman gauge --  the double logarithm $\ln^2\lambda^2$ is fully determined by the crossed-box diagram shown in the right panel of Fig.~\ref{fig:treeandnlo}. More precisely, the double-logarithmic contributions to the planar box and the vertex diagrams cancel in their sum in this particular gauge.

NNLO virtual corrections to muon-electron scattering are known in the massless approximation for more than two decades~\cite{Bern:2000ie}, and they have been computed more recently for a massive muon and a massless electron in~\cite{Bonciani:2021okt}. Partial NNLO results for non-vanishing muon \emph{and} electron masses are also available~\cite{CarloniCalame:2020yoz,Banerjee:2020rww}, but they do not contain the contributions from the two-loop box diagrams yet. 
We are therefore not in the position to extract the leading logarithms in the equal-mass limit at NNLO from the existing results in the literature.

\subsection{The double-logarithmic amplitude}
\label{subsec:DLamplitude}

We now briefly discuss the traditional approach for resumming the double-logarithmic contributions to the backward-scattering amplitude to all orders following~\cite{Gorshkov:1966qd,Berestetskii:1982qgu}. Before we turn to two- and higher-loop contributions, we will explain the simplifications that arise in the double-logarithmic approximation at one-loop order in some detail. In Feynman gauge, we just saw that the double logarithms are entirely encoded in the crossed-box diagram shown in the right panel of Fig.~\ref{fig:treeandnlo}. In the equal-mass limit we thus start from
\begin{align}
\label{eq:diagd}
    i{\cal M} &\simeq (4\pi\alem)^2 \int \!\frac{d^d k}{(2\pi)^d} \frac{\big[\bar{u}^{(e)}(\bar{p}) \gamma^\mu (\slashed{k}+m) \gamma^\nu u^{(e)}(p)\big] \,\big[\bar{u}^{(\mu)}(p) \gamma_\nu (\slashed{k}+m) \gamma_\mu u^{(\mu)}(\bar{p}) \big]}{(k^2-m^2)^2 (k-p)^2 (k-\bar{p})^2} \,,
\end{align}
where from now on we use the symbol $\simeq$ to indicate the double-logarithmic approximation. These double logarithms arise from the kinematic configuration in which the virtual lepton propagators go on-shell, which fixes their virtuality to $\mathcal{O}(\lambda^2)$, whereas both light-cone components $n_-k$ and $n_+k$ are small of $\mathcal{O}(\lambda)$. The loop momentum thus scales as a soft momentum in the SCET terminology of Sec.~\ref{sec:SCET}. Expanding the numerator in $\lambda$ shows that only the perpendicular components of all Dirac matrices contribute,
\begin{align}
 &\big[\bar{u}^{(e)}(\bar{p}) \gamma^\mu (\slashed{k}+m) \gamma^\nu u^{(e)}(p)\big] \,\big[\bar{u}^{(\mu)}(p) \gamma_\nu (\slashed{k}+m) \gamma_\mu u^{(\mu)}(\bar{p}) \big]
\nonumber \\ 
 & = \, \big[\bar{u}^{(e)}_{\bar{\xi}} \gamma_\perp^\mu (\slashed{k}_\perp+m) \gamma_\perp^\nu u^{(e)}_\xi\big] \, \big[\bar{u}^{(\mu)}_\xi \gamma_{\perp \nu} (\slashed{k}_\perp+m) \gamma_{\perp \mu} u^{(\mu)}_{\bar{\xi}}\big] + \mathcal{O}(\lambda) \,.
\end{align}
Due to the vanishing perpendicular components of the external momenta, we can further substitute this by
\begin{align}
    \frac{k_\perp^2}{d-2} \, \big[\bar{u}^{(e)}_{\bar{\xi}} \gamma_\perp^\mu \gamma_\perp^\alpha \gamma_\perp^\nu u^{(e)}_\xi\big] \,\big[\bar{u}^{(\mu)}_\xi \gamma_{\perp \nu} \gamma_{\perp\alpha} \gamma_{\perp \mu} u^{(\mu)}_{\bar{\xi}}\big] + m^2 \big[\bar{u}^{(e)}_{\bar{\xi}} \gamma_\perp^\mu \gamma_\perp^\nu u^{(e)}_\xi\big] \,\big[\bar{u}^{(\mu)}_\xi \gamma_{\perp \nu} \gamma_{\perp \mu} u^{(\mu)}_{\bar{\xi}}\big] \,.
\end{align}
In $d=4$ dimensions the first term reduces to the tree-level Dirac structure, and it hence contributes to the form factor $F_1(\lambda)$. The second term, on the other hand, only gives a contribution to the helicity-flipping form factor $F_2(\lambda)$, which is single-logarithmic and can thus be neglected. After some algebra we find that the double-logarithmic contribution to $F_1(\lambda)$ is entirely contained in the integral
\begin{align}
\label{eq:oneloopscalarint}
 F_1^{(1)}(\lambda) &\simeq 16i\pi^2 s \int \!\frac{d^d k}{(2\pi)^d} \frac{k_\perp^2 }{(k^2-m^2)^2 (k-p)^2 (k-\bar{p})^2}
 \\
&= 16i\pi^2 s  \int \!\frac{d^d k}{(2\pi)^d} \bigg\{
\frac{1}{(k^2-m^2) (k-p)^2 (k-\bar{p})^2} 
+  \frac{m^2-(n_-k)(n_+k)}{(k^2-m^2)^2 (k-p)^2 (k-\bar{p})^2}\bigg\} \,.
\nonumber
\end{align}
In the following we concentrate on the first term, which contains only one massive lepton propagator, and by repeating the same analysis for the second term, one can show that it does not produce any double logarithms.

As already stated, we consider the loop momentum to be soft, and in this approximation the photon propagators become eikonal,
\begin{align}
    (k-p)^2 +i0 \approx -\sqrt{s} \, n_-k +i0\,, \qquad\qquad
    (k-\bar{p})^2 +i0 \approx -\sqrt{s} \, n_+k +i0\,.
\end{align}
We then integrate over the perpendicular components and examine the analytic structure of the resulting $(n_-k)$-integral in the complex plane. One finds that the form factor is given by the discontinuity associated with the lepton propagator with
\begin{align}
\label{eq:DLF1NLO}
 F_1^{(1)}(\lambda) &\simeq \frac{(4\pi)^\eps}{\Gamma(1-\eps)} \int_0^\infty \frac{d(n_-k)}{n_-k} \int_0^\infty \frac{d(n_+k)}{n_+k} \, \theta\big((n_+k)(n_-k)-m^2\big) \big((n_+k)(n_-k)-m^2\big)^{-\eps} \,.
\end{align}
In the effective-field-theory formulation the longitudinal integrations would be unconstrained in the soft region. In the traditional approach, however, one imposes hard cutoffs on the longitu\-dinal integrations,
\begin{align}
 \frac{m^2}{\sqrt{s}}\,\ll\, n_+k \,\ll\, \sqrt{s} \,, \qquad\qquad
 \frac{m^2}{\sqrt{s}}\,\ll\, n_-k \,\ll\, \sqrt{s}\,,
\end{align}
such that one can perform the integral in $d=4$ dimensions. In terms of the dimensionless variables $x=n_+k/\sqrt{s}$ and $y=n_-k/\sqrt{s}$ this gives
\begin{align}
 F_1^{(1)}(\lambda) &\simeq
\int_{\lambda^2}^{1} \frac{dx}{x} \int_{\lambda^2/x}^{1} \frac{dy}{y}
= \frac{1}{2} \ln^2\lambda^2 \,,
\end{align}
in agreement with~\eqref{eq:F1oneloop}.

\begin{figure}
\centering
\includegraphics[scale=2.3]{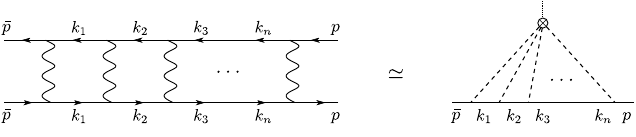}
	\caption{Ladder-type diagrams that determine the double-logarithmic contribution to the scattering amplitude in Feynman gauge. Notice again that the upper line is twisted. The double logarithms arise from the kinematic configuration in which each lepton propagator goes on-shell and the photon propagators become eikonal, see the discussion around \eqref{eq:eikonal}. At each order in perturbation theory the Dirac algebra can then be simplified such that the double logarithms are contained in a simpler scalar integral (right graph), where solid (dashed) lines indicate massive (massless) propagators.}
	\label{fig:ladder}
\end{figure}

At higher orders the dominant corrections $\sim\alem^{n} \ln^{2n} m/\sqrt{s}$ can be computed in a similar fashion. The key observation is that they purely arise from ladder-type photon exchanges (in the twisted notation, see Fig.~\ref{fig:ladder}), where the loop integrations within each individual sub-diagram generate double logarithms in the same way as discussed for the one-loop case above. This implies that one only has to consider momentum configurations in which all lepton propagators go on-shell and the photon propagators become eikonal, 
\begin{align}
\label{eq:eikonal}
    (k_{i}-k_{i-1})^2 +i0\approx -(n_+k_i)(n_-k_{i-1}) +i0\,.
\end{align}
%
This corresponds to the situation in which the longitudinal components of the lepton momenta are strongly ordered (see also \cite{Gorshkov:1966qd,Berestetskii:1982qgu}),
\begin{align}
\label{eq:strongordering}
 \frac{m^2}{\sqrt{s}}&\approx n_+\bar p \,\ll\, n_+k_1 \,\ll\,  \ldots \,\ll\, n_+k_n \,\ll\, n_+ p \approx \sqrt{s}\,,\nonumber\\
 \frac{m^2}{\sqrt{s}}&\approx n_- p \,\ll\, n_-k_n \,\ll\, \ldots \,\ll\, n_- k_1\,\ll\, n_-\bar p \approx \sqrt{s} \,.
\end{align}
Focusing again on the double-logarithmic contribution, one can simplify the numerator as before to reduce the ladder diagram with $n$ rungs to a simpler scalar integral,
\begin{align}
\label{eq:nlooptriangle}
 F_1^{(n)}(\lambda) \simeq \, &(-16i\pi^2)^n (-s) \int \!\frac{d^dk_1}{(2\pi)^d} \dots \frac{d^dk_n}{(2\pi)^d} \;\frac{1}{k_1^2-m^2}\,\dots\, \frac{1}{k_n^2-m^2}
\nonumber \\
 &
 \times\, \frac{1}{(\bar{p}-k_1)^2} \,\frac{1}{(k_1-k_2)^2} \,\dots\, 
\frac{1}{(k_{n-1}-k_n)^2} \,\frac{1}{(k_n-p)^2} \,,
\end{align}
which in fact also holds at tree level with $F_1^{(0)}(\lambda) = (-s)/(p-\bar{p})^2 = 1 + \mathcal{O}(\lambda)$. The double logarithms of the backward-scattering amplitude are thus entirely contained in a simpler set of scalar integrals, depicted in Fig.~\ref{fig:ladder}, which provides a well-defined template for studying the problem of endpoint singularities in SCET by means of a method-of-regions analysis. In fact, although in a different physical context, precisely these integrals were studied before in~\cite{Boer:2018mgl}.  

Due to the eikonal structure \eqref{eq:eikonal} of the photon propagators, the integrals over the perpendicular components can again be performed trivially, which yields a contribution associated with the discontinuity of each lepton propagator. In terms of the dimensionless variables $x_i=n_+k_i/\sqrt{s}$ and $y_i=n_-k_i/\sqrt{s}$, the resulting factors $\theta\big((n_+k_i)(n_-k_i)-m^2\big)$ furthermore combine with the phase-space constraints from \eqref{eq:strongordering} to the following representation
\begin{align}
\label{eq:nestedintegrals}
 F_1^{(n)}(\lambda) \simeq &\int \frac{dx_1}{x_1} \int \frac{dy_1}{y_1} \,\dots\, \int \frac{dx_n}{x_n} \int \frac{dy_n}{y_n} \; 
\theta(x_1 y_1-\lambda^2)\,\dots\, \theta(x_n y_n -\lambda^2) \nonumber \\
  & \times\,
	\theta(1-y_1) \, \theta(y_1-y_2) \, \dots \, \theta(y_n-\lambda^2) \; 
	\theta(1-x_n) \, \theta(x_n-x_{n-1}) \, \dots \, \theta(x_1-\lambda^2) \nonumber \\
  = &\int_{\lambda^2}^1 \frac{dx_1}{x_1} \int^{1}_{x_1} \frac{dx_2}{x_2} \, \dots \, \int^{1}_{x_{n-1}} \frac{dx_n}{x_n} 
  \; \int^1_{\lambda^2/x_1} \frac{dy_1}{y_1} \int^{y_1}_{\lambda^2/x_2} \frac{dy_2}{y_2} \, \dots\, \int^{y_{n-1}}_{\lambda^2/x_n} \frac{dy_n}{y_n} \,.
\end{align}
For any given $n$, these integrals can be performed explicitly, which yields the simple result
\begin{align}
 F_1^{(n)}(\lambda)\simeq \frac{1}{n!(n+1)!} \; \ln^{2n}\lambda^2 \,.
\end{align}
Therefore we obtain for the resummed expression \begin{align}
\label{eq:BesselI}
 F_1(\lambda) = \sum_{n=0}^\infty \left(\frac{\alem}{2\pi}\right)^n F_1^{(n)}(\lambda) \simeq \sum_{n=0}^\infty \,\frac{\left(\frac{\alem}{2\pi} \ln^2\lambda^2\right)^n}{n!(n+1)!} = \frac{I_1\left(2\sqrt{\frac{\alem}{2\pi}\ln^2\lambda^2}\right)}{\sqrt{\frac{\alem}{2\pi}\ln^2\lambda^2}} \,,
\end{align}
with the modified Bessel function of the first kind $I_1(2\sqrt{z})$. More rigorously, this result can be obtained by introducing an additional cutoff parameter for the integrals~\eqref{eq:nestedintegrals} as discussed in~\cite{Berestetskii:1982qgu}. The so-defined auxiliary function then obeys an implicit integral equation which can be solved in Laplace space. The result~\eqref{eq:BesselI} has an interesting asymptotic behaviour in the limit $z \to \infty$,
\begin{align}
    \frac{I_1(2\sqrt{z})}{\sqrt{z}} \approx \frac{1}{2\sqrt{\pi}} \frac{e^{2\sqrt{z}}}{z^{3/4}}\,,
\end{align}
and the form factor thus grows exponentially in $|\ln \lambda^2|$, contrary to the exponential fall off from the standard Sudakov factor in the same approximation.


\section{Formulation in SCET}
\label{sec:SCET}

We now address the question how the resummed expression~\eqref{eq:BesselI} in the double-logarithmic approximation is reproduced in the effective field theory. To do so, we derive a ``bare'' factorization theorem for the backward-scattering amplitude, which, as we shall see, suffers from endpoint-divergent convolution integrals. These endpoint divergences spoil the standard approach for resumming the logarithmic contributions to all orders using renormalization-group techniques. We will come back to this point in the following section, and focus here on the details of the derivation of the factorization theorem.

As is familiar from other examples that involve massive fermions, the loop integrals require a rapidity regulator in the effective theory to make the contributions from individual momentum regions well-defined. The relevant effective theory is called \scettwo, and we will follow a two-step matching procedure in this section, in which one first integrates out hard fluctuations associated with the scale $\sqrt{s}\gg m$. The degrees of freedom in the intermediate effective theory \scetone~are hard-collinear, anti-hard-collinear and soft modes with respective scaling in the power-counting parameter $\lambda$:
\begin{align}
    \text{hard:} \qquad k^\mu &\sim (1,1,1)\sqrt{s}\,, \nonumber \\
    \text{hard-collinear:} \qquad k^\mu &\sim (1,\sqrt{\lambda},\lambda) \sqrt{s}\,, \nonumber \\
    \text{anti-hard-collinear:} \qquad k^\mu &\sim (\lambda,\sqrt{\lambda},1) \sqrt{s}\,, \qquad\qquad\nonumber \\
    \text{soft} \qquad k^\mu &\sim (\lambda,\lambda,\lambda) \sqrt{s} \,.
\end{align}
In \scetone~the large component of a hard-collinear fermion field scales as $\lambda^{1/2}$, a soft fermion field as $\lambda^{3/2}$, and photon fields scale as their corresponding momentum components. Interactions between the (anti-)hard-collinear and soft sectors can be eliminated from the \scetone~Lagrangian by a decoupling transformation.

The on-shell amplitude contains, however, collinear and anti-collinear external states with virtualities $k^2= \mathcal{O}(m^2)$ much smaller than the hard-collinear scale $m \sqrt{s}$. As anti\-cipated in \eqref{eq:externalmom}, their generic scaling behaviour is given by
\begin{align}
    \text{collinear:} \qquad k^\mu &\sim (1,\lambda,\lambda^2) \sqrt{s}\,, \nonumber \\
    \text{anti-collinear:} \qquad k^\mu &\sim (\lambda^2,\lambda,1) \sqrt{s} \,.
\end{align}
In a second matching step, one therefore integrates out (anti-)hard-collinear fluctuations, and the resulting effective theory \scettwo~contains collinear, anti-collinear and soft modes. The large component of a collinear fermion field now scales as $\lambda$, and interactions between the (anti-)collinear and soft sectors are always highly off-shell and therefore not part of the \scettwo~Lagrangian.

\subsection{Method-of-regions analysis} 
\label{subsec:MoR}

In order to illustrate some subtleties associated with the regularization prescription, we find it instructive to analyze the momentum regions of the one-loop scalar integral in the first term of the second line in~\eqref{eq:oneloopscalarint}, which captures the double-logarithmic contributions at this order, in some detail. We define 
\begin{align}
\label{eq:onelooptrianglefull}
{\cal I} &\equiv 16i\pi^2 s \,\tilde{\mu}^{2\eps}
 \int \!\frac{d^d k}{(2\pi)^d} \,
\frac{1}{(k^2-m^2) (k-p)^2 (k-\bar{p})^2} = \frac12 \ln^2\lambda^2 + \frac{2\pi^2}{3} + \mathcal{O}(\lambda) \,,
\end{align}
where the usual $i0$-prescription of the Feynman propagators is understood. The integral is IR- and UV-finite, but for later purposes we expressed the integral measure in $d = 4-2\eps$ dimensions with $\tilde{\mu}^2=\mu^2 e^{\gamma_E}/(4\pi)$. 

The integral receives contributions from four momentum regions: the loop momentum $k^\mu$, which is assigned to the massive lepton propagator, can either become hard, collinear, anti-collinear, or soft. It is well known that individual momentum regions of loop integrals with massive propagators can be ill-defined, even when they are evaluated in $d$ dimensions. The reason is that dimensional regularization is attached to the transverse momentum components in light-cone coordinates, and in the presence of mass terms the regularization does not carry over to the longitudinal components after integrating over the transverse space~\cite{Becher:2011dz}.
In other words, dimensional regularization as a boost-invariant regularization scheme only distinguishes momentum modes by their virtuality, but in \scettwo~all three long-distance modes have equal virtuality, $k^2= \mathcal{O}(m^2)$, and are separated only by a parametrically large boost. To consistently regularize the appearing singularities then requires the use of rapidity regulators. Many different prescriptions have been proposed in the literature (see e.g.~\cite{Chiu:2009yx,Becher:2011dz,Chiu:2012ir,Echevarria:2015byo,Li:2016axz,Ebert:2018gsn}), and in the following we use an analytic regulator that supplements each integral measure with a factor
\begin{align} \label{eq:symmetricregulator}
    d^d k \;\to\; d^d k \,\Big(\frac{\nu}{2k_0-i0}\Big)^\reg\,,
\end{align}
with $2k_0 = n_+k + n_-k$. This regulator preserves the $n_-^\mu \leftrightarrow n_+^\mu$ symmetry of the process, and the $i0$-prescription is chosen to be consistent with the analytic structure of the eikonal propagators, as we will see below. To this end it is important that the loop momentum is always assigned to the massive lepton propagator in the direction of fermion flow, as is illustrated by the diagram in Fig.~\ref{fig:ladder}. Such an adhoc prescription, of course, violates gauge invariance, which is only restored once the contributions from the soft, collinear and anti-collinear regions are combined. A discussion about a gauge-invariant implementation of a rapidity regulator is beyond the scope of the present paper, and it is in fact not needed to resum the double-logarithmic contributions as we will see in Sec.~\ref{sec:resummation}.

With the analytic regulator in place, the contributions from the individual momentum regions can readily be computed. As the analytic regulator is supposed to regularize rapidity divergences only, it is important that the limit $\reg\to0$ is taken before the dimensional regulator is set to zero. We then obtain the following contributions to the integral in \eqref{eq:onelooptrianglefull}:

\paragraph{Hard region:} The contribution from the hard region simply corresponds to the massless integral, which can be performed in $d$ dimensions:
\begin{align}
\label{eq:hardregion}
 {\cal I}^{\text{(hard)}} &= 16i\pi^2s \,\tilde{\mu}^{2\eps} 
\int \!\frac{d^d k}{(2\pi)^d} \,\frac{1}{k^2 (k-p)^2 (k-\bar{p})^2} \nonumber \\
&= \frac{1}{\eps^2} + \frac{1}{\eps}\ln\frac{\mu^2}{s} + \frac12 \ln^2\frac{\mu^2}{s}-\frac{\pi^2}{12} + \mathcal{O}(\eps) \,.
\end{align}
By dropping the fermion masses, we observe that the integral has become IR-divergent, and these divergences are fully controlled by the dimensional regulator.

\paragraph{Collinear region:} 
In the collinear region one of the photon propagators becomes eikonal, and the rapidity regulator simplifies according to $2k_0 = n_+k + \mathcal{O}(\lambda^2)$: 
\begin{align}
 {\cal I}^{\text{(anti-col)}} &= -16i\pi^2 \sqrt{s} \,\tilde{\mu}^{2\eps} \!\int \!\frac{d^d k}{(2\pi)^d} \,\Big( \frac{\nu}{n_+k-i0} \Big)^\reg 
\frac{1}{(k^2-m^2+i0) (k^2-2kp+m^2 +i0)(n_+k-i0)}  \nonumber \\
 &= e^{\eps \gamma_E} \,\Gamma(\eps) \Big(\frac{\mu^2}{m^2}\Big)^\eps \Big(\frac{\nu}{\sqrt{s}}\Big)^\reg \int_0^1 dx \; x^{-1-\reg} \, (1-x)^{-2\eps} \nonumber \\ 
 &= -\frac{1}{\reg \eps} - \frac{1}{\reg}\ln\frac{\mu^2}{m^2} - \frac{1}{\eps}\ln\frac{\nu}{\sqrt{s}} - \ln\frac{\mu^2}{m^2}\ln\frac{\nu}{\sqrt{s}} + \frac{\pi^2}{3} +  \mathcal{O}(\reg,\eps) \,.
\end{align}
To evaluate the integral we pick up the residues in the small momentum component $n_-k$. The integral over the perpendicular components then acquires a UV divergence $1/\eps$. The analytic structure furthermore restricts the integration domain of the remaining integral over the large component $n_+ k = \sqrt{s}\, x$ to the interval $x \in [0,1]$. This integral is endpoint divergent as $x\to0$ and generates a $1/\reg$~pole in the rapidity regulator. As expected, the collinear mode is characterized by a virtuality $\mu^2\sim m^2$ and a large energy $\nu \sim \sqrt{s}$.

\paragraph{Anti-collinear region:} As our choice of the analytic regulator preserves the $n_-^\mu \leftrightarrow n_+^\mu$ symmetry, the anti-collinear region simply gives the identical result:
\begin{align}
 {\cal I}^{\text{(anti-col)}} = {\cal I}^{\text{(col)}}\,.
\end{align}

\paragraph{Soft region:} In the soft region both photon propagators become eikonal, but now with hard-collinear scaling $(k-p)^2 \sim (k-\bar{p})^2\sim m \sqrt{s}$. Since $n_-k \sim n_+k$ in this region, the $i0$-prescription of the analytic regulator becomes relevant, which, as argued before, has been chosen to be in line with the $i0$-prescription of the eikonal propagators. We find
\begin{align}
\hspace{-0.5mm} {\cal I}^{\text{(soft)}} &= 16i\pi^2 \,\tilde{\mu}^{2\eps} \int \!\frac{d^d k}{(2\pi)^d} \Big( \frac{\nu}{n_+k+n_-k-i0} \Big)^\reg \frac{1}{(k^2-m^2+i0)(n_-k-i0)(n_+k-i0)}   \\
 &= \frac{e^{\eps \gamma_E} \mu^{2\eps}}{\Gamma(1-\eps)} 
\int_0^\infty d\xi \; \frac{\xi^{-\eps}}{\xi+m^2} \int_0^\infty \frac{d(n_+k)}{n_+k} \Big( \frac{\nu \, n_+k}{\xi+m^2+(n_+k)^2}\Big)^\reg \nonumber \\
 &= \frac{2}{\reg\eps} + \frac{2}{\reg}\ln\frac{\mu^2}{m^2} + \frac{2}{\eps}\ln\frac{\nu}{m} + 2\ln\frac{\mu^2}{m^2}\ln\frac{\nu}{m}  - \frac{1}{\eps^2} - \frac{1}{\eps}\ln\frac{\mu^2}{m^2} - \frac12 \ln^2\frac{\mu^2}{m^2} + \frac{\pi^2}{12} + \mathcal{O}(\reg,\eps) \,.
 \nonumber
\end{align}
Here we used the residue theorem in the variable $n_-k$. The analytic structure then fixes the variable $n_+k$ to the entire positive real axis, and the rapidity regulator mixes the longitudinal $n_+k$ with the $k_\perp$ integration (we wrote $\xi \equiv -k_\perp^2>0$). The latter again generates a UV singularity  $1/\eps$, whereas the integral over $n_+k$ acquires a $1/\reg$ singularity from both limits $n_+k \to \infty$ and $n_+k \to 0$. These rapidity divergences reflect the overlap with the collinear and anti-collinear region for fixed invariant mass $(n_-k)(n_+k) \sim m^2$, respectively. From the last line we read off that the soft region is characterized by a virtuality $\mu^2 \sim m^2$ and a small energy $\nu \sim m$. 

Given the results from all regions, one easily verifies that their sum reproduces the full integral in~\eqref{eq:onelooptrianglefull}. For the following discussion it is, moreover, instructive to contrast the calculation with a second one that uses the alternative rapidity regulator $\nu^\reg/(n_+k-i0)^\reg$. Obviously, this leaves the hard as well as the collinear region unaffected, whereas the soft region becomes scaleless in this scheme, as the regulator no longer mixes the longitudinal and the perpendicular integrations. To compensate this mismatch, the anti-collinear region then gives the sum of the former soft and anti-collinear regions with
\begin{align}
 {\cal I}^{\text{(anti-col)}^\prime} = &\frac{1}{\reg\eps} + \frac{1}{\reg} \ln\frac{\mu^2}{m^2} + \frac{1}{\eps} \ln\frac{\nu\sqrt{s}}{m^2} + \ln\frac{\mu^2}{m^2}\ln\frac{\nu \sqrt{s}}{m^2} 
 - \frac{1}{\eps^2} - \frac{1}{\eps}\ln\frac{\mu^2}{m^2}- \frac12\ln^2\frac{\mu^2}{m^2} + \frac{5\pi^2}{12} \,,
\end{align}
which shows that $\mu^2 \sim m^2$ in this region, whereas the small component $n_+k$ now scales with $\nu\sim m^2/\sqrt{s}$. While we will mainly work with the symmetric regulator from \eqref{eq:symmetricregulator} in this section, we will exploit the very fact that the soft integrals are scaleless in the alternative regularization scheme to resum the double-logarithmic corrections in Sec.~\ref{sec:resummation}.

\subsection{Bare factorization theorem}
\label{sec:factorization}

We now formulate the problem in the effective theory and derive an all-order factorization theorem for the scattering amplitude using a two-step QED~$\to$~\scetone~$\to$~\scettwo~matching procedure. This factorization must be understood among \emph{bare} objects that are not yet expanded in the dimensional regulator $\eps$, and implicitly supplemented by a suitable rapidity regulator $\reg$. 

\begin{figure}
\centering
\includegraphics[scale=4.0]{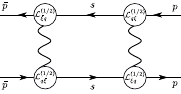}
	\caption{Exchange of two virtual soft leptons at the one-loop level.}
	\label{fig:softleptons}
\end{figure}

To keep the discussion concise, we focus in this section on the helicity-conserving form factor $F_1(\lambda)$. We will find that the respective bare factorization theorem at leading power in $\lambda$ consists of two terms that can be written in the form
\begin{align}
\label{eq:barefactheorem}
 F_1(\lambda) \,=\,& \int \frac{dx}{x} \int \frac{dy}{y} \; H(xys) 
\left\{f_c(x) \,f_{\bar{c}}(y) +\widetilde{f}_c(x)\,\widetilde{f}_{\bar{c}}(y) \right\} 
\\
& + \int \frac{dx}{x} \int \frac{dy}{y} \int \frac{d\rho}{\rho} \int \frac{d\omega}{\omega} \; J_{hc}(\rho x\sqrt{s})\; J_{\bar hc}(\omega y\sqrt{s}) 
\nonumber\\
& \quad
\times \left[S(\rho,\omega) \left\{f_c(x) \,f_{\bar{c}}(y) +\widetilde{f}_c(x)\,\widetilde{f}_{\bar{c}}(y) \right\}  + \widetilde{S}(\rho,\omega) \left\{f_c(x) \,\widetilde{f}_{\bar{c}}(y) +\widetilde{f}_c(x)\,f_{\bar{c}}(y) \right\}  \right]\,.\nonumber
\end{align}
Up to an overall sign, the respective expression for the form factor $F_2(\lambda)$ takes precisely the same form, but with one or three helicity-flipping functions (indicated by the tilde), instead of zero or two, as in the above expression for $F_1(\lambda)$. The term in the first line arises from integrating out hard modes, and it starts at tree level with the hard-photon exchange shown in the left panel of Fig.~\ref{fig:treeandnlo}. The second term in the following two lines, depicted in Fig.~\ref{fig:softleptons}, only starts at one-loop order and involves the exchange of soft leptons, which in SCET are induced by power-suppressed currents. We now address each of these structures in turn.

\paragraph{First term:} The first term reflects the standard hard-scattering picture. It arises from integrating out hard fluctuations at a scale $\sqrt{s}\gg m$, and involves a hard-coefficient function that is convoluted with forward matrix elements of non-local SCET-2 operators. As the hard-collinear scale is irrelevant for this contribution, the matching from the intermediate theory \mbox{\scetone}~onto the final theory \scettwo~is trivial in this case. The relevant operator takes the form 
\begin{align}
\label{eq:O1}
    \mathcal{O}_{1}(\sigma,\tau) = &
		\big[\bar{\chi}^{(e)}_{\bar{c}}(0) \frac{\slashed{n}_-}{2} P_R  \, \chi_{\bar{c}}^{(\mu)}(\sigma n_-)\big] \,
		\big[\bar{\chi}^{(\mu)}_c(\tau n_+) \frac{\slashed{n}_+}{2} P_R  \, \chi^{(e)}_c(0) \big] 
		 + (R\to L) \,,
\end{align}
whose Dirac structure can be directly associated with the tree-level expressions in~\eqref{eq:LOlp} and~\eqref{eq:Fierz}, and the lepton field operators $\chi_c=W^\dagger_c \xi_c = W^\dagger_c (\slashed{n}_-\slashed{n}_+/4)\psi_c$ are the usual gauge-invariant SCET building blocks, i.e.~the large components of the spinor fields in each collinear sector dressed with collinear Wilson lines.

As the collinear and anti-collinear sectors are decoupled in \scettwo, one can study flavour-off-diagonal forward matrix elements of this operator separately in each sector. We then define the two collinear functions $f_c(x)$ and $\widetilde{f}_c(x)$ occurring in the factorization theorem \eqref{eq:barefactheorem} via
\begin{align}
\label{eq:phicdef}
    &\bra{\mu^-(p)} \,\bar{\chi}_c^{(\mu)}(\tau n_+) \frac{\slashed{n}_+}{2} 
		P_{R(L)} \,\chi^{(e)}_c(0) \,\ket{e^-(p)} \nonumber \\
    &\qquad =\int \!dx\, e^{ix\tau n_+p} \left\{f_c(x) \, \big[\bar{u}_\xi^{(\mu)} \frac{\slashed{n}_+}{2} P_{R(L)} u_\xi^{(e)}\big] + \widetilde{f}_c(x) \, \big[\bar{u}_\xi^{(\mu)} \frac{\slashed{n}_+}{2} P_{L(R)} u_\xi^{(e)}\big] \right\} \,,
\end{align}
and in complete analogy for the anti-collinear functions $f_{\bar{c}}(y)$ and $\widetilde{f}_{\bar{c}}(y)$. Since the fermion mass contributes at leading power in \scettwo, the matrix element must be parametrized in terms of two functions in each sector, a helicity-conserving function $f_c(x)$ and a helicity-flipping function $\widetilde{f}_c(x)$. These functions can be considered as a particular manifestation of generalized parton distributions (GPDs), see e.g.\ \cite{Radyushkin:1997ki,Ji:1998pc,Diehl:2003ny} for reviews. In our case the GPDs are off-diagonal in lepton flavour, but diagonal in external momenta (in the language of GPDs this corresponds to the limit of zero skewness).

We emphasize that the collinear functions describe both lepton and anti-lepton distributions, and their support is given by the interval $x \in [-1,1]$, where $x$ measures the fraction of the collinear lepton momentum that enters the hard interaction, relative to the external momentum $p$ (the momentum fraction $x$ is in fact carried by \emph{both} lepton flavours, as a consequence of translation invariance).\footnote{For $x\geq 0$ one considers the situation that an electron with momentum fraction $x$ inside the initial-state electron is absorbed at the origin, and a muon with the same momentum fraction is emitted back at the point $\tau n_+$ into the final-state muon. For $x\leq 0$ one considers the situation that a
positron with momentum fraction $(-x)$ 
is emitted at the origin into the final state (where it recombines with an initial-state electron), and an anti-muon with the same momentum fraction (generated by pair creation from the initial state) is absorbed at the point $\tau n_+$.} As the anti-lepton distribution in a lepton only starts at ${\cal O}(\alem^2)$, it is not relevant for the resummation of the double logarithms that we envisage in Sec.~\ref{sec:resummation}. We can therefore restrict the longitudinal integrations to the interval $[0,1]$ for this discussion later on. 

The hard function $H(xys)$ in the first term of \eqref{eq:barefactheorem} can  be determined in a QED~$\to$ \mbox{\scetone} matching calculation. Despite matching a $2\to2$ scattering amplitude, the specific kinematics of the process fixes the argument of the hard function to a single variable $xys$. In practice the matching is most conveniently performed on-shell in dimensional regularization, setting all IR scales to zero along with $p^\mu \to x \sqrt{s} \,n_-^\mu/2$ and $\bar{p}^\mu \to y \sqrt{s} \,n_+^\mu/2$. When working in dimensional regularization, the operator basis needs to be extended by evanescent operators, but once again this complication is irrelevant for the resummation of the leading double logarithms and can therefore be disregarded in our analysis in Sec.~\ref{sec:resummation}.

As usual in an effective-field-theory treatment, the IR divergences of the hard function must match the UV divergences of the collinear functions defined above. But since the considered scattering amplitude is IR-divergent by itself -- see~\eqref{eq:F1oneloop} -- the hard function contains an additional source of IR divergences that is not captured by those functions. We already discussed the physics of these IR divergences in Sec.~\ref{subsec:fixed-order}; they are associated with another form of soft momentum modes that do not scale with $m\ll\sqrt{s}$, but rather with the experimental resolution energy $E_\gamma\ll m$. As the corresponding soft function is scaleless to all orders in perturbation theory in our setup, we refrain from including it in the factorization theorem \eqref{eq:barefactheorem} for clarity. We nevertheless explicitly show in App.~\ref{app:imagpart} that the UV divergences of this soft function match the remnant IR singularities of the hard function at one-loop order, as they should in a consistent effective-theory description.

The helicity structure of the process furthermore implies that the factorization theorem for the form factor $F_1(\lambda)$ contains an even number of helicity-flipping functions, whereas the one for the form factor $F_2(\lambda)$ contains an odd number. The hard function is, moreover, normalized to $H(xys)=1$ at tree level in our conventions, and the helicity-conserving functions evaluate to $f_c(x) = \delta(1-x)$ and $f_{\bar{c}}(y) = \delta(1-y)$ at this order, whereas the helicity-flipping ones are only non-zero at the loop level. The first term in the factorization theorem \eqref{eq:barefactheorem} therefore reproduces the tree-level result $F_1(\lambda)=1$.

\paragraph{Second term:} 

The second contribution to the factorization theorem \eqref{eq:barefactheorem} is more subtle. It arises from the exchange of two virtual soft leptons and starts only at the one-loop level, as illustrated in Fig.~\ref{fig:softleptons}.  Here, for example, the collinear electron in the initial state turns into an anti-collinear electron not through the exchange of a hard photon, but rather through two consecutive exchanges of a hard-collinear and an anti-hard-collinear photon, allowing the virtual electron to become soft.
We note that other one-loop graphs that involve soft lepton configurations are power-suppressed, and only the particular ladder-type diagram with two soft-lepton propagators contributes at leading power. 

In \scetone~the soft contribution to the scattering amplitude can be written as the matrix element of a time-ordered product that involves subleading Lagrangian insertions
\begin{align}
\label{eq:Lxiqs}
    {\cal L}_{\xi q}^{(1/2)}(x) &= \bar{\chi}^{(\ell)}_{hc}(x) \,\slashed{\cal A}^\perp_{hc}(x) \,\psi^{(\ell)}_s(x_-) \,,
\end{align}
which couples a soft lepton to a hard-collinear lepton and photon. Here 
\begin{align}
 {\cal A}^{\perp,\mu}_{hc} = e\left[ A^{\perp,\mu}_{hc} - \frac{i \partial^{\perp,\mu} n_+ A_{hc}}{i n_+ \partial} \right]   
\end{align}
denotes the usual gauge-invariant building block for perpendicular hard-collinear photon fields. For a consistent expansion in the power-counting parameter $\lambda$, the soft fields must be multipole expanded in interactions with hard-collinear fields, which is the reason why the soft lepton field only depends on $x_-^\mu \equiv (n_+ x) n_-^\mu/2$. Similar expressions with $x_+^\mu \equiv (n_- x) n_+^\mu/2$ hold for interactions of soft fields with anti-hard-collinear fields. We will see, however, shortly that the multipole expansion is subtle for the considered process, because of the specific kinematic configurations of backward scattering.

In order to describe the soft contribution to the scattering amplitude we thus start from the \scetone~operator
\begin{align}
    \mathcal{O}_{2} &= 
		T\,\bigg\{i\,{\cal L}_{\bar\xi q}^{(1/2)}(0),i\!\int\! d^dx_1 \,{\cal L}_{q\bar\xi }^{(1/2)}(x_1),i\!\int\! d^dx_2 \,{\cal L}_{\xi q}^{(1/2)}(x_2),i\!\int\! d^dx_3 \,{\cal L}_{q\xi }^{(1/2)}(x_3)\bigg\} \,,
\end{align}
which after decoupling the soft photons from the hard-collinear and anti-hard-collinear fields with suitable field redefinitions factorizes into
\begin{align}
\label{eq:O2inSCET1}
    \mathcal{O}_{2} &= \int\! d^dx_1 \int\! d^dx_2 \int\! d^dx_3 \;\;
		T\,\Big\{ \big[ \bar{\chi}^{(e)}_{\bar hc } \,\slashed{\cal A}^\perp_{\bar hc} \big]^{\alpha}(0)\,\big[ \slashed{\cal A}^\perp_{\bar hc} \chi^{(\mu)}_{\bar hc} \big]^{\beta}(x_1)\Big\}
				\\
				&\times\;
		T\,\Big\{ \big[ \bar{\chi}^{(\mu)}_{hc} \,\slashed{\cal A}^\perp_{hc} \big]^{\gamma}(x_2+x_3)\,\big[ \slashed{\cal A}^\perp_{hc} \chi^{(e)}_{hc}  \big]^{\delta}(x_3)\Big\}
				\nonumber\\
&\times\;
		T\,\Big\{ \big[  S_{n_+}^\dagger \psi^{(e)}_s \big]^{\alpha}(0)
		\,
		\big[ \bar\psi^{(\mu)}_s \bar S_{n_+} \big]^{\beta}(x_{1+})
		\,
		\big[ S_{n_-}^\dagger \psi^{(\mu)}_s \big]^{\gamma}(x_{2-}+x_3)
		\,
		\big[ \bar\psi^{(e)}_s \bar S_{n_-} \big]^{\delta}(x_{3})\Big\}
		\,,
		\nonumber
		\end{align}
where we have made the spinor indices $(\alpha,\beta,\gamma,\delta)$ explicit. We furthermore shifted one of the integration variables, and  distinguished between soft Wilson lines associated with incoming particles (denoted by $\bar S_n$) and outgoing particles (denoted by $S_n^\dagger$). Their precise definition can be found in App.~\ref{app:imagpart}. Notice that despite appearance all the soft fields in~\eqref{eq:O2inSCET1} have been multipole-expanded, where the reference point for the multipole expansion is $x=0$ in the anti-hard-collinear direction but $x = x_3$ in the hard-collinear direction. As the relevant forward matrix element of the hard-collinear operator in the second line is translation invariant, one can shift the hard-collinear fields to the spacetime points $x_2$ and $0$, respectively, to obtain a structure that is similar to the one in the first line. The coordinates $x_1$ and $x_2$ therefore scale inversely to an anti-hard-collinear and a hard-collinear momentum, respectively, whereas the coordinate $x_3$ scales inversely to a soft momentum. This is a consequence of the specific kinematics of backward scattering, in which both virtual leptons in Fig.~\ref{fig:softleptons} can only become soft simultaneously when the difference of the two (anti-) hard-collinear external momenta scales as a soft momentum.\footnote{%
A different situation arises in the limit in which all kinematic invariants 
are large compared to the electron mass, i.e. $s,|t|,|u|\gg m^2$. 
In that case, as has been shown in \cite{Becher:2007cu} for Bhabha scattering, large logarithms 
in the ratio of the electron mass and the center-of-mass energy  can be \emph{multiplicatively}
factorized using SCET methods in a straight-forward manner.
}

It may seem surprising that an operator with four insertions of a power-suppressed interaction can give a leading-power contribution, in particular since the operator $\mathcal{O}_1$ in \eqref{eq:O1} only involves leading-twist projections. By applying the \scetone~power-counting rules described at the beginning of this section, one finds that $\mathcal{O}_1\sim\lambda^2$, whereas $\mathcal{O}_2\sim\lambda^2$ precisely since the spacetime point $x_3$ scales inversely to a soft momentum, which yields the necessary enhancement. In the following we refer to this phenomenon as the \emph{soft-enhancement mechanism}.

In the next step one integrates out the hard-collinear and anti-hard-collinear fluctuations and matches the intermediate theory \scetone~onto the final theory \scettwo, in which the collinear, soft and anti-collinear sectors are decoupled. The corresponding matching relation for the forward matrix element of the hard-collinear operator in the second line of \eqref{eq:O2inSCET1} reads
\begin{align}
\label{eq:SCET2matchingsecondterm}
		&\bra{\mu^-(p)} \,T\, \big[ \bar{\chi}^{(\mu)}_{hc} \,\slashed{\cal A}^\perp_{hc} \big]^{\gamma}(x_2)\,\big[ \slashed{\cal A}^\perp_{hc} \chi^{(e)}_{hc}  \big]^{\delta}(0) \,\ket{e^-(p)}
		= 2ie^2 \int \!\frac{d^d k}{(2\pi)^d} \; e^{i k x_2} \!
		\int \! dx \, \frac{J_{hc}\big((n_-k)x(n_+p)\big)}{- (n_-k)x(n_+p)} 
		\nonumber\\
		&\quad \times
		\left\{f_c(x) \, \big[\bar{u}_\xi^{(\mu)} \frac{\slashed{n}_+}{2} P_{R} u_\xi^{(e)}\big] \, \big[ \frac{\slashed{n}_-}{2} P_{R} \big]^{\delta\gamma}+ \widetilde{f}_c(x) \, \big[\bar{u}_\xi^{(\mu)} \frac{\slashed{n}_+}{2} P_{L} u_\xi^{(e)}\big] \, \big[ \frac{\slashed{n}_-}{2} P_{R} \big]^{\delta\gamma}  + (R\leftrightarrow L) \right\}\,,
\end{align}
where we rearranged the collinear spinors with the help of a Fierz trans\-formation. The matching relation shows that the second term in the factorization theorem can be expressed in terms of the same collinear functions $f_c(x)$ and $\widetilde{f}_c(x)$ defined in \eqref{eq:phicdef}, which are, however, in this case convoluted with a non-trivial hard-collinear matching coefficient. This jet function, which we normalized to $J_{hc}\big(x(n_+p)(n_-k)\big)=1$ at tree level, only depends on the light-cone component $n_-k\sim\lambda$. In complete analogy one then parametrizes the forward matrix element of the anti-hard-collinear operator.

We finally turn to the vacuum matrix element of the soft operator in the last line of \eqref{eq:O2inSCET1}. The most general parametrization of this matrix element with four open spinor indices is rather involved. After contraction with the hard-collinear matrix elements, however, only two soft functions contribute to the backward-scattering amplitude at leading power. One is helicity conserving,
\begin{align}
\label{eq:S}
& S(\rho,\omega) = {-i\frac{e^2}{8\pi^2} }
\int \! d(n_-x_1) \int \! d(n_+x_2)   
		\;\, e^{+\frac{i}{2} \rho (n_+ x_{2})} 
		\; e^{-\frac{i}{2} \omega (n_-x_{1})}
		\int \!d^d x_3
		\\
& \,
 \times \bra{0} \,
		T \big[ \bar\psi^{(\mu)}_s \bar S_{n_+} \big](x_{1+})
		\frac{\slashed{n}_+}{2} P_{R(L)}
		\big[  S_{n_+}^\dagger \psi^{(e)}_s \big](0)
		\big[ \bar\psi^{(e)}_s \bar S_{n_-} \big](x_{3})
		\frac{\slashed{n}_-}{2} P_{R(L)}
		\big[ S_{n_-}^\dagger \psi^{(\mu)}_s \big](x_{2-}+x_3) 
		\ket{0},
		\nonumber
\end{align}
and one is helicity flipping,
\begin{align}
\label{eq:Stilde}
& \widetilde{S}(\rho,\omega) = {-i\frac{e^2}{8\pi^2} }
\int \! d(n_-x_1) \int \! d(n_+x_2)   
		\;\, e^{+\frac{i}{2} \rho (n_+ x_{2})} 
		\; e^{-\frac{i}{2} \omega (n_-x_{1})}
		\int \!d^d x_3
		\\
& \,
 \times \bra{0} \,
		T \big[ \bar\psi^{(\mu)}_s \bar S_{n_+} \big](x_{1+})
		\frac{\slashed{n}_+}{2} P_{R(L)}
		\big[  S_{n_+}^\dagger \psi^{(e)}_s \big](0)
		\big[ \bar\psi^{(e)}_s \bar S_{n_-} \big](x_{3})
		\frac{\slashed{n}_-}{2} P_{L(R)}
		\big[ S_{n_-}^\dagger \psi^{(\mu)}_s \big](x_{2-}+x_3) 
		\ket{0}.
		\nonumber
\end{align}
One may think that the soft functions can only depend on the product $\rho \omega$ because of boost invariance, but the rapidity regulator that is needed in endpoint-divergent convolutions to make the soft integrals well-defined breaks this symmetry, and they should therefore be considered as functions of two independent variables (see also~\eqref{eq:soft:treelevel} for the explicit expressions of the soft functions at leading order). As the soft contribution to the form factors only starts at $\mathcal{O}(\alem)$, we furthermore found it convenient to absorb a factor $e^2$ into the definition of the soft functions.

Our results for the soft contribution to the form factors can finally be expressed in terms of a four-fold convolution over the variables $(x,y,\rho,\omega)$. When performing these convolutions, one has to carefully restore the $i0$-prescriptions in the various denominators, which yields the discontinuity of the soft functions, as we will show explicitly in the one-loop calculation below. Starting at the two-loop level, the soft functions are in fact not IR-finite, because of soft singularities associated with modes of energy $E_\gamma\ll m$. As these IR-divergences are expected to factorize from the entire scattering amplitude, they must also be present in the second term of the factorization theorem. We finally remark that -- with minor modifications in the definitions of the collinear and soft functions -- the bare factorization formula also holds for non-equal lepton masses or for a sufficiently small but non-zero scattering angle.

\subsection{Endpoint-divergent convolutions}
\label{subsec:endpoint}

Although the relevant scales in the process seem to be completely disentangled in the bare factorization theorem \eqref{eq:barefactheorem}, it cannot be used to resum the logarithmic corrections to all orders with renormalization-group techniques. As we will explicitly show in this section by computing all ingredients that are required at the NLO level, the various convolutions give rise to divergences in the dimensional regulator $\eps$ and the rapidity \mbox{regulator $\reg$}, which spoil the renormalization program. While our prior interest consists in making this problem explicit, we will also clarify in the following which ingredients are needed in the double-logarithmic approximation that we will address in detail in Sec.~\ref{sec:resummation}. Throughout this section, we employ the notation we introduced earlier, writing $F_1(\lambda) = \sum_{n=0}^\infty \left(\frac{\alem}{2\pi}\right)^n F_1^{(n)}(\lambda)$ etc.

\paragraph{First term:} 
We start with the hard contribution to the form factor, which evaluates to 
$F_1^{(0)}(\lambda)=1$ at tree level. As we have seen in the method-of-regions analysis in Sec.~\ref{subsec:MoR}, the IR divergences of the hard loops are regularized in dimensional regularization, and we may therefore perform the corresponding QED~$\to$~\mbox{\scetone} matching calculation without an additional rapidity regulator. Since the hard function is a single-scale object, radiative corrections can then only generate powers of $(xys)^{-\eps}$ at each loop order, which should not be expanded in the dimensional regulator, since the convolution integrals are -- as we will see shortly -- in general endpoint-divergent in the limits $x \to 0$ and $y\to 0$. By properly factoring out this piece, but expanding the remaining functions \mbox{in $\eps$}, we find that the one-loop contribution to the bare hard-coefficient function is given by
\begin{align}
\label{eq:hardfun}
 H^{(1)}(xys) = \Big(\frac{\mu^2}{x y s}\Big)^\eps \left[ \frac{1}{\eps^2} + \frac{2\pi i}{\eps} - \frac{1}{\eps}  - \frac{13\pi^2}{12} - 4 - \frac{2 n_\ell}{3\eps} - \frac{10 n_\ell}{9}\right] \,, 
\end{align}
which refers to a specific definition of evanescent operators, but the first two terms in the square bracket are scheme-independent. Similar to the full-theory calculation discussed in Sec.~\ref{subsec:fixed-order}, the double pole in $\eps$ can again be extracted purely from the ladder-type diagram in Feynman gauge.

\begin{figure}
\centering
\includegraphics[scale=2.2]{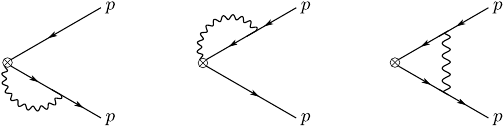}
	\caption{Radiative corrections to the collinear matrix element~\eqref{eq:phicdef} at $\mathcal{O}(\alem)$.}
	\label{fig:1loopcollME}
\end{figure}

We next consider the collinear matrix elements, defined in~\eqref{eq:phicdef}, which evaluate to $f_c^{(0)}(x)=\delta(1-x)$ and $\widetilde{f}_c^{(0)}(x)=0$ at tree level. As the tree-level distributions vanish at the endpoint $x\to 0$, their convolution with the hard function is endpoint-finite to all orders in perturbation theory. We will show next, however, that the one-loop matrix elements do not vanish as $x\to 0$, which leads to endpoint-divergent convolution integrals. Specifically, the endpoint divergences manifest themselves in the form $\int_0^1 dx \,x^{-1-n \eps-\alpha}$ in the convolution with the $n$-loop hard function, i.e.~they produce a pole in the dimensional regulator $\eps$ at $n$ loops, but they require the rapidity regulator $\reg$ for the convolution with the tree-level hard function for $n=0$.

The relevant Feynman diagrams for the calculation of the collinear functions $f_c(x)$ and $\widetilde{f}_c(x)$ at $\mathcal{O}(\alem)$ are shown in Fig.~\ref{fig:1loopcollME}. In our calculation we implement the rapidity regulator in the symmetric form \eqref{eq:symmetricregulator}, which ensures that the corresponding anti-collinear functions are given by the same expressions with $x\to y$ and $n_+p\to n_-\bar p$. Accounting for the renormalization factors of the lepton fields, we find
\begin{align}
\label{eq:1loopbarefc}
    f_c^{(1)}(x) &= \Big( \frac{\mu^2 e^{\gamma_E}}{m^2}\Big)^\eps \Big( \frac{\nu}{n_+p} \Big)^\alpha\,\Gamma(\eps) \bigg\{\delta(1-x) \left( \frac{1}{\eps} + \frac12 + \mathcal{O}(\eps,\alpha) \right) \nonumber\\
		& \qquad \qquad \qquad \qquad \qquad \qquad 
    + \theta(x)\theta(1-x) \, (1 - \eps) \,
		\frac{1 + x^2 - \eps (1 - x)^2}{x^\alpha(1 - x)^{1 + 2 \eps}} \bigg\}\,, \nonumber \\
    \widetilde{f}_c^{(1)}(x) &= 
		\Big(\frac{\mu^2 e^{\gamma_E}}{m^2}\Big)^\eps  
			\Big(\frac{\nu}{n_+p} \Big)^\alpha \,\theta(x)\theta(1-x) \, \Gamma(1+\eps)\, (1 - \eps) \,x^{-\alpha}
		(1 - x)^{1 - 2 \eps} \,,
\end{align}
where we kept the exact dependence on both regulators $\eps$ and $\reg$, except in the delta-function term, whose convolution with the hard function is uncritical. In the limit $x\to 0$, we thus observe that the collinear functions do not vanish anymore,
\begin{align}
\label{eq:1loopbarefcendpoint}
    f_c^{(1)}(x\to 0) &= 	\Big(\frac{\mu^2 e^{\gamma_E}}{m^2}\Big)^\eps  
			\Big(\frac{\nu}{n_+p} \Big)^\alpha 
			\, \Gamma(\eps) \,
		(1 - \eps)^2 \,x^{-\alpha}\,,
		\nonumber\\
    \widetilde{f}_c^{(1)}(x\to 0) &= \Big(\frac{\mu^2 e^{\gamma_E}}{m^2}\Big)^\eps  
			\Big(\frac{\nu}{n_+p} \Big)^\alpha 
			\, \Gamma(1+\eps)\, (1 - \eps) \,x^{-\alpha} \,,
\end{align}
which produces endpoint divergences in the convolution with the hard function as argued above. Interestingly, the endpoint behaviour of the collinear functions is -- in Feynman gauge -- again entirely captured by the ``ladder-type'' diagram in the right panel of Fig.~\ref{fig:1loopcollME}, whereas the contributions from the Wilson-line diagrams vanish linearly as $x\to 0$. Notice also that the helicity-conserving function $f_c(x)$ is UV-divergent, and it hence contributes in the double-logarithmic approximation, whereas the helicity-flipping function $\widetilde{f}_c(x)$ is not.

We already emphasized that one should not expand the bare collinear functions in the various regulators before performing the convolution with the hard function, but let us for the moment nevertheless do so to cross-check our results. Setting the rapidity regulator to zero, and expanding the expression \eqref{eq:1loopbarefc} in plus-distributions,
we obtain
\begin{align}
    f_c^{(1)}(x) &= \frac{1}{\eps} \left[\frac{1+x^2}{1-x}\right]_+ + \mathcal{O}(\eps^0)\,,
\end{align}
and we thus recover the one-loop quark-to-quark splitting kernel, as expected for a forward GPD. In contrast to standard applications of GPDs, we consider here, however, moments that are divergent in the limit $x\to 0$, which prevents one from renormalizing these objects in the standard sense.

\paragraph{Second term:} 
As the soft contribution to the form factor starts as $\mathcal{O}(\alem)$, we  need to evaluate the respective ingredients only at tree level to the considered NLO accuracy. Starting from the definitions in~\eqref{eq:S} and~\eqref{eq:Stilde}, we find for the tree-level soft functions 
\begin{align}
\label{eq:soft:treelevel}
    S^{(1)}(\rho,\omega) &= \frac{1}{2\pi i} \,\Big(\frac{\mu^2e^{\gamma_E}}{m^2-\rho\omega-i0}\Big)^\eps  \,\Big(\frac{\nu}{\rho+\omega-i0} \Big)^\alpha \; \Gamma(\eps)\,(1-\eps)\,, 
		\nonumber \\
    \widetilde{S}^{(1)}(\rho,\omega) &= \frac{1}{2\pi i} \,\Big(\frac{\mu^2e^{\gamma_E}}{m^2-\rho\omega-i0}\Big)^\eps  \,\Big(\frac{\nu}{\rho+\omega-i0} \Big)^\alpha \;  \Gamma(1+\eps) \; \frac{m^2}{m^2-\rho\omega-i0}\,.
\end{align}
With the explicit results at hand, we can now illustrate how the discontinuities of the soft functions arise in the convolution integrals. To this end, we restore the \mbox{$i0$-prescription} in the tree-level hard-collinear propagators, which yields
\begin{align}
\label{eq:disc}
&\int \frac{d\rho}{\rho-i0} \int \frac{d\omega}{\omega-i0}  \;S^{(1)}(\rho,\omega) 
 \nonumber\\
&\qquad
= \frac{1-\eps}{\Gamma(1-\eps)}\,\int_0^\infty \frac{d\rho}{\rho} \int_0^\infty \frac{d\omega}{\omega}  \; \theta(\rho\omega-m^2) \left(\frac{\mu^2 e^{\gamma_E}}{\rho\omega-m^2}\right)^\eps 
\left(\frac{\nu}{\rho+\omega}\right)^\reg\,,
\end{align}
where we have used that the soft function has a branch cut in the lower half-plane of the complex variable $\rho$ only for $\omega>0$. We emphasize that the analytic properties of the chosen rapidity regulator in \eqref{eq:symmetricregulator} are crucial at this step. We thus obtain a non-zero contribution to the double con\-vo\-lu\-tion from the discontinuity $$\text{Disc} \, (m^2-\rho\omega)^{-\eps} = (m^2-\rho\omega-i0)^{-\eps}-(m^2-\rho\omega+i0)^{-\eps} \,, $$ 
which implies $\rho\omega > m^2$, but in general radiative corrections will also induce a second branch with $\rho\omega < m^2$ and $\rho,\omega >0$ (see App.~\ref{app:nestedintegrals}).

The convolution integrals are most easily evaluated by splitting the phase space into two regions with $\rho>\omega$ and $\rho<\omega$. As the rapidity regulator \eqref{eq:symmetricregulator} preserves the $n_-^\mu \leftrightarrow n_+^\mu$ symmetry, both regions yield identical contributions, and we  furthermore find it convenient to introduce the variables $u=\rho\omega$ and $v=\rho+\omega$ to simplify the subsequent calculation. We then find that both of these integrations are endpoint-divergent; the limit $u\to\infty$ produces a $1/\eps$ pole and the limit $v\to\infty$ a $1/\alpha$ pole. We also note in passing that the soft integrals would be scaleless if we had chosen the alternative regulator $\nu^\reg/(n_+k-i0)^\reg$, but in this case the calculation of the anti-collinear functions would have been more complicated. This is similar to what we have seen in the method-of-regions analysis in Sec.~\ref{subsec:MoR}.

\paragraph{The double-logarithmic approximation:} 
We now have assembled all pieces to illustrate how the double logarithms are generated at the one-loop order in the effective-theory formulation. Our calculation, of course, should not only reproduce the double logarithms, but the full NLO result from \eqref{eq:F1oneloop}, but to prove this would require a more sophisticated treatment of evanescent operators, which we do not pursue here.

Starting from the bare factorization theorem \eqref{eq:barefactheorem}, the one-loop contribution to the form factor $F_1(\lambda)$ becomes
\begin{align}
\label{eq:NLO:doublelog}
 F_1^{(1)}(\lambda) &=  H^{(1)}(s) + 2 \int \frac{dx}{x} \,f_c^{(1)}(x)
+ \int \frac{d\rho}{\rho-i0} \int \frac{d\omega}{\omega-i0}  \;S^{(1)}(\rho,\omega) 
\\
&\simeq \frac{1}{\eps^2} \Big(\frac{\mu^2}{s}\Big)^\eps
- \frac{2}{\reg\eps} \Big(\frac{\mu^2}{m^2}\Big)^\eps \Big(\frac{\nu}{\sqrt{s}}\Big)^\reg
+ \bigg\{ \frac{2}{\reg\eps} - \frac{1}{\eps^2} \bigg\} 
\Big(\frac{\mu^2}{m^2}\Big)^\eps \Big(\frac{\nu}{m}\Big)^\reg
= \frac12 \ln^2\lambda^2\,,
\nonumber
\end{align}
where we have approximated $(n_+p) = (n_-\bar{p}) \approx \sqrt{s}$ at leading power. We thus reproduce the double-logarithmic correction of the NLO result in \eqref{eq:F1oneloop}, but is interesting to note how the double logarithms are generated in the effective theory from a peculiar interplay between the various regions. While the pattern is the same as for the scalar integral that we discussed in Sec.~\ref{subsec:MoR}, we appreciate now that they do so via endpoint-divergent convolution integrals.

While there seems to be no gain in rederiving the one-loop logarithms via the effective-field-theory formulation, it will be critical for resumming the logarithmic corrections to all orders, as we will show in the following section.
In particular, the double-logarithmic approximation -- which for this particular example is known for more than 50 years -- provides in our opinion a prime example for studying the physics of endpoint singularities in SCET. On the double-logarithmic level, the factorization theorem \eqref{eq:barefactheorem} in fact simplifies tremendously, since one can immediately drop all helicity-flipping functions, constrain the integration boundaries of the variables $x$ and $y$ to the unit interval, and disregard the contributions from evanescent operators. If in addition we switch to the alternative regulator $\nu^\reg/(n_+k-i0)^\reg$ that makes all moments of the soft function scaleless to all orders, we see that the double logarithms are captured by a surprisingly simple and minimal structure,
\begin{align}
\label{eq:fac:minimal}
 F_1(\lambda) \,\simeq\,& \int_0^1 \frac{dx}{x} \int_0^1 \frac{dy}{y} \; H(xys) 
\, f_c(x) \,f_{\bar{c}}(y)  \,,
\end{align}
where the anti-collinear function is, of course, no longer given by the expressions from above, due to the different choice of the rapidity regulator. We thus obtain a factorization theorem that involves two leading-twist collinear functions that are convoluted with a hard-coefficient function that depends only on the product of the convolution variables. The structure looks completely harmless, since all interactions and all twist projections are performed at leading power. Nevertheless, and somewhat against the common lore, the convolutions are endpoint-singular in the limits $x\to0$ and $y\to0$ in a complicated way. 

We believe that the key ingredient, which distinguishes the current process from other examples with endpoint divergences in SCET that were studied recently (see e.g.~\cite{Liu:2019oav,Beneke:2020ibj,Beneke:2022obx}), is the two-fold convolution in the hard-scattering term, which contains products of endpoint-divergent moments. We will perform a detailed comparison of the present QED process with the bottom-induced $h\to \gamma\gamma$ decay in Sec.~\ref{sec:hgg}, but we think that the structure we are seeing in \eqref{eq:fac:minimal} is more general, and applies 
in a similar fashion to more complicated hard-exclusive processes in QCD, like exclusive heavy-to-light $B$-meson decays, which involve endpoint-divergent convolutions of both heavy- and light-meson light-cone distribution amplitudes with a perturbative hard-scattering kernel. In contrast to these examples, however, the current QED backward-scattering process is, of course, much simpler and it is in some sense a well-defined perturbative playground for studying non-trivial aspects of soft-collinear factorization.

\section{Resummation of double logarithms}
\label{sec:resummation}

In the previous section we saw that the convolution integrals in the factorization theorem \eqref{eq:barefactheorem} are divergent in the limits $x,y\to 0$ and $\rho,\omega\to\infty$ at NLO, which prevents one from renormalizing the bare quantities prior to performing the convolutions. 
In particular, the explicit calculation in \eqref{eq:NLO:doublelog} illustrates that the convolutions in the first term produce a $1/\reg$ pole that cancels against a similar contribution from the convolutions in the second term. For this cancellation to happen, the two terms must describe the same physics in the endpoint region of the respective convolutions. In this section we will show that the collinear functions can indeed be refactorized in the limit $x\to 0$ in terms of the very same jet and soft functions that appear in the second term. While this idea of \emph{endpoint refactorization} seems to be crucial for dealing with endpoint-divergent convolutions in SCET~\cite{Liu:2019oav,Beneke:2020ibj,Beneke:2022obx}, it was first introduced by one of us in~\cite{Boer:2018mgl} in the context of exclusive $B$-meson decays. The refactorization conditions in conjunction with the simplistic structure of the factorization theorem \eqref{eq:fac:minimal} in the presence of the asymmetric rapidity regulator furthermore allow us to resum the double-logarithmic corrections to all orders using consistency relations. Interestingly, we will find that this resummation that yields the well-known Bessel function from \eqref{eq:BesselI} proceeds via an infinite tower of collinear-anomaly exponents.

\subsection{Refactorization of collinear matrix elements}
\label{subsec:refactorization}

Endpoint-divergent moments of the collinear functions $f_c(x)$ and $\widetilde{f}_c(x)$ signal a sensitivity to unnaturally small light-cone components of the collinear momenta, which invalidates their power counting $n_+ k = x \sqrt{s}=\mathcal{O}(\sqrt{s})$. As argued before, the physics in the endpoint region must reflect the asymptotic contribution of the second term in the factorization theorem, and one should therefore count $x=\mathcal{O}(\lambda)$ in the endpoint region, while keeping the virtuality $k^2=\mathcal{O}(m^2)$ fixed. In other words, the backward-scattering amplitude receives two different types of contributions: one for generic $x=\mathcal{O}(1)$ for which the standard hard-scattering picture applies, and one for $x=\mathcal{O}(\lambda)$ for which the collinear functions become multi-scale objects and must be refactorized.

\begin{figure}
\centering
\includegraphics[scale=2.0]{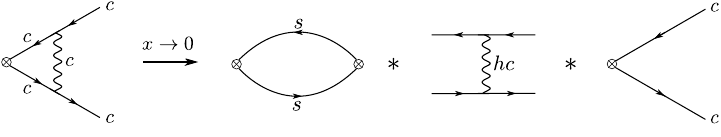}
	\caption{Refactorization of the endpoint contribution to the collinear functions at $\mathcal{O}(\alem)$.} 
	\label{fig:refactorization}
\end{figure}

In the explicit NLO calculation from Sec.~\ref{subsec:endpoint}, we found that the dominant endpoint contribution to the collinear functions is entirely captured in Feynman gauge by the last diagram in Fig.~\ref{fig:1loopcollME}. In the endpoint region with $x=\mathcal{O}(\lambda)$, this diagram can be refactorized in the form illustrated in Fig.~\ref{fig:refactorization}. The goal of the present section consists in deriving the precise form of this refactorization condition, and in proving that the same jet and soft functions that were defined in Sec.~\ref{sec:factorization} arise in this context. To this end, we proceed in close analogy to the discussion of the soft contribution to the factorization theorem from the previous section. Specifically, we reconsider the collinear matrix element in the limit $x\to 0$ in \mbox{\scetone}, where the endpoint configuration is produced via subleading Lagrangian insertions, similar to the mechanism shown in Fig.~\ref{fig:softleptons},
\begin{align}
    &\int \!d\tau\, e^{-ix\tau n_+p}  \;
		\bra{\mu^-(p)} \,\bar{\chi}_c^{(\mu)}(\tau n_+) \frac{\slashed{n}_+}{2} 
		P_{R} \,\chi^{(e)}_c(0) \,\ket{e^-(p)} 
		 \;\xrightarrow{\scriptstyle\rule[-2pt]{0pt}{2pt}{x\to 0}}\;
		\int \!d\tau\, e^{-ix\tau n_+p}  \;
		\\
		& 		\bra{\mu^-(p)} \,
		T\bigg\{\!\big[ \bar\psi^{(\mu)}_s \bar S_{n_+} \big](\tau n_+) 
		\frac{\slashed{n}_+}{2} P_{R}  \big[  S_{n_+}^\dagger \psi^{(e)}_s \big](0),
		i\!\!\int\! d^dx_1 \,{\cal L}_{q\xi }^{(1/2)}(x_1) , i\!\!\int\! d^dx_2 \,{\cal L}_{\xi q}^{(1/2)}(x_2) \!\bigg\}
		\ket{e^-(p)}.
		\nonumber
\end{align}
By following exactly the same steps that were used in the previous section, one finds that the endpoint contribution to the collinear functions can be refactorized in the form\footnote{There is a subtle point hidden in our notation, namely the soft functions inherit the rapidity regulator that has been expanded for $n_+k\gg n_-k$ in the collinear region, and it therefore does not correspond to the one that is implied by the definition of the soft functions in \eqref{eq:S} and \eqref{eq:Stilde}. But apart from this point, one obtains exactly the same functions and we decided not to introduce a new notation for these objects.}
\begin{align}
 f_c(x\to 0) \,=\,& 
 \int \frac{dx'}{x'}  \int \frac{d\rho}{\rho} \; 
J_{hc}\big(\rho x'(n_+p)\big) \left[f_c(x') \, S\big(\rho,x(n_+p)\big)
+ \widetilde{f}_c(x') \, \widetilde{S}\big(\rho,x(n_+p)\big)\right]\,,
\nonumber\\
 \widetilde{f}_c(x\to 0) \,=\,& 
 \int \frac{dx'}{x'}  \int \frac{d\rho}{\rho} \; 
J_{hc}\big(\rho x'(n_+p)\big) \left[\widetilde{f}_c(x') \, S\big(\rho,x(n_+p)\big) 
+ f_c(x') \, \widetilde{S}\big(\rho,x(n_+p)\big)\right]\,,
\label{eq:refactorization}
\end{align}
which holds for the bare functions, up to contributions which vanish at least linearly in the limit $x\to 0$. We recall that a Fierz transformation was applied in the derivations in Sec.~\ref{sec:factorization}, and the refactorization conditions are therefore in general modified by evanescent operators, which we do not include in this work. Analogous expressions can be derived for the anti-collinear functions $f_{\bar{c}}(y)$ and $\widetilde{f}_{\bar{c}}(y)$ in the limit $y\to 0$.

The above refactorization conditions, which are a key result of our paper, have an interesting structure. First of all, they can be diagonalized by switching from the helicity basis to the vector- and axial-vector currents. More importantly, one observes that the same collinear functions that are being refactorized in the limit $x\to 0$ appear on the right-hand side of these equations within a two-fold convolution, and the relations should therefore be considered as implicit formulae for the collinear functions. As the soft functions start as $\mathcal{O}(\alem)$, the collinear functions on the right-hand side enter, however, at lower order in perturbation theory than on the left-hand side of the equations. But even worse, the convolutions in the variables $x'$ and $\rho$ turn out to be endpoint-divergent -- they in fact involve the same divergent moments of $f_c(x')$ -- and the various functions must therefore again be supplemented by the rapidity regulator. The problem of endpoint-divergent convolutions thus repeats itself on the level of the refactorization conditions. We will see below that the consistency of the effective-theory formulation requires this iterative structure.

Focusing again on the double-logarithmic contributions, the refactorization conditions take a simpler form,
\begin{align}
\label{eq:refactorization:DL}
 f_c(x\to 0) \,\simeq\,& 
 \int \frac{dx'}{x'}  \; f_c(x') \int \frac{d\rho}{\rho} \; 
J_{hc}\big(\rho x'(n_+p)\big) \,S\big(\rho,x(n_+p)\big)\,,
\end{align}
which is the one that was anticipated in~\cite{Boer:2018mgl}. Subtracting the asymptotic limit $x\to0$ from the collinear function $f_c(x)$ thus amounts to a \emph{convolution},
\begin{align}
 f_c(x) - f_c(x\to 0) \,\simeq\,& 
 \int \frac{dx'}{x'}  \; f_c(x') \bigg\{ x\, \delta(x-x') - \int \frac{d\rho}{\rho} \; 
J_{hc}\big(\rho x'(n_+p)\big) \,S\big(\rho,x(n_+p)\big)\bigg\}\,.
\end{align}
This relation can formally be solved for $f_c(x)$, which leads to an infinite sum of nested convolutions of the soft and jet functions as discussed in~\cite{Boer:2018mgl}. While such a representation may not be particularly illuminating, the structure of the refactorization condition~\eqref{eq:refactorization:DL} has several important implications:
\begin{itemize}
\item
On dimensional grounds, the jet function can only depend on powers of $\big(\rho x'(n_+p)\big)^{-\eps}$ at each loop order, similar to what we have seen for the hard function in Sec.~\ref{subsec:endpoint}. As the soft function is a boost-invariant object, it can only depend on the product of the arguments $\rho x(n_+p)$ (modulo terms associated with the rapidity regulator, which are irrelevant for this aspect). One can therefore rescale the convolution variable $\rho$ in a way to move the $x$-dependence of the soft function into the jet function. The upshot is that the collinear functions receive \emph{positive} powers of $x^\eps$ through hard-collinear loops in the refactorization formula, starting at two loops. Hence, in the convolution with the $n$-loop hard function, the moment 
\begin{align}
\label{eq:collinear:moment}
\big\langle x^{-1-n\eps} \big\rangle_{f_c}
\equiv \int_0^1 dx \,x^{-1-n \eps} \,f_c(x)
\end{align}  
generates rapidity divergences \emph{for every $n$}, since the positive powers of $x^\eps$ inherent in $f_c(x)$ can cancel the $x^{-n \eps}$ at a certain order. More precisely, the one-loop collinear function produces a rapidity divergence in the convolution with the tree-level hard function as we have seen explicitly in Sec.~\ref{subsec:endpoint}, the two-loop collinear function does so in the convolution with the one-loop and tree-level hard function, etc. The very structure that \emph{all} moments $\langle x^{-1-n\eps} \rangle_{f_c}$ generate rapidity divergences starting at $\mathcal{O}(\alem^{n+1})$ will be important for the resummation of the double-logarithmic contributions in the following section.
\item
As the convolutions in $x'$ and $\rho$ are in general endpoint-divergent, both functions $f_c(x')$ and $S\big(\rho,x(n_+p)\big)$ must be evaluated with the rapidity regulator $\alpha$ in place. We argued in Sec.~\ref{sec:factorization}, however, that the collinear functions are particular manifestations of GPDs, which are, of course, finite in the limit $\alpha\to 0$. In other words, \emph{moments} of the collinear functions suffer from rapidity divergences, but not the collinear functions themselves. This in turn implies that the rapidity divergences associated with the $x'$- and $\rho$-convolutions must cancel against each other within the refactorization formula. By the usual collinear-anomaly argument~\cite{Becher:2010tm,Becher:2011pf} this cancellation then generates a large rapidity logarithm,    
\begin{align}
\frac{1}{\alpha} + \ln \Big(\frac{\nu}{n_+p}\Big) 
- \frac{1}{\alpha} - \ln \Big(\frac{\nu}{x(n_+p)}\Big) 
= \ln x\,.
\end{align}
We thus expect that the collinear functions have a logarithmic dependence on $x$ in the endpoint region, even for finite values of the dimensional regulator, i.e.~when terms like $x^\eps$ are \emph{not} expanded in the limit $\eps\to0$. These logarithms first show up at the two-loop order, and they are borne out by an explicit two-loop calculation of the scalar integrals~\eqref{eq:nlooptriangle} in~\cite{Boer:2018mgl}.
\end{itemize}
While these comments are somewhat technical, they should become clearer in the following sections, and they are also illustrated by an explicit ``higher-loop'' calculation in App.~\ref{app:nestedintegrals}. We close this section with a remark that may have confused the careful reader. Namely one may wonder if the soft functions in the refactorization formulae can be made scaleless by a suitable choice of the rapidity regulator, and in this way the collinear functions would simply vanish at the endpoint $x\to 0$ at each loop order. This is, unfortunately, not possible since the soft functions themselves are not scaleless, see e.g.~\eqref{eq:soft:treelevel}, but only their \emph{double} moments if the asymmetric regulator is applied. The refactorization conditions in contrast only involve single moments of the soft functions, which are not scaleless in any regularization scheme.

\subsection{Resummation using consistency relations}
\label{subsec:anomalies}

In this section we will show how the double logarithms that we derived earlier in Sec.~\ref{subsec:DLamplitude} with diagrammatic methods are reproduced in the effective theory. As the factorization theorem \eqref{eq:barefactheorem} only holds for bare quantities, we cannot use renormalization-group techniques for this purpose. Instead we will exploit the simplified version of the factorization theorem \eqref{eq:fac:minimal}, which captures the full information on the double-logarithmic contributions in a scheme with an asymmetric rapidity regulator $\nu^\reg/(n_+k-i0)^\reg$. We will then resum the double-logarithmic contributions using three ingredients: (i) consistency, which requires that the poles in the various regulators cancel on the level of the form factor; (ii) scale separation, which ties the characteristic logarithms for each function to these poles; (iii) the refactorization condition \eqref{eq:refactorization:DL}, which provides another non-trivial input as we will see below.

On the double-logarithmic level one can safely neglect all effects associated with the running of the coupling, and we can hence work with a vanishing beta function. This implies that in $d=4-2\eps$ spacetime dimensions the dimensionless coupling $\alem(\mu)\equiv \alem$ depends on the scale $\mu$ only trivially through $d\alem/d\ln\mu = - 2\eps \alem$. We then rewrite the bare factorization theorem in the form
\begin{align}
\label{eq:DLfacscales}
 F_1(\lambda) \,\simeq\,& \int_0^1 \frac{dx}{x} \; 
f_c\Big(x;\frac{\mu}{m},\frac{\nu}{\sqrt{s}}\Big)
\int_0^1 \frac{dy}{y} \; 
f_{\bar{c}}\Big(y;\frac{\mu}{m},\frac{\nu\sqrt{s}}{m^2}\Big) \; 
H\Big(\frac{\mu^2}{xys}\Big)\,,
\end{align}
where we have made the dependence on all dimensionless scale ratios explicit. As we work with bare unexpanded quantities, this dependence is of the form $(\mu/m)^\eps$ and $(\nu/\sqrt{s})^\reg$ etc, see e.g.~\eqref{eq:hardfun} and \eqref{eq:1loopbarefc} for explicit expressions of the one-loop hard and collinear functions. We re-emphasize that these are still the bare quantities, which formally do not depend on the renormalization scale $\mu$, but in our notation we want to track the logarithms associated with each function that are generated once the expansion in the various regulators is performed.

We now insert the perturbative expansion of the hard function
\begin{align}
H\Big(\frac{\mu^2}{xys}\Big) &=
\sum_{n=0}^\infty \,\left(\frac{\alem}{2\pi}\right)^n 
\Big(\frac{\mu^2}{x y s}\Big)^{n\eps} \;
\bigg\{\frac{h^{(n)}}{\eps^{2n}} 
+ \mathcal{O}\Big(\frac{1}{\eps^{2n-1}}\Big)\bigg\}\,,
\end{align}
where $h^{(0)}=1$ and we only kept the highest pole in $\eps$ at each loop order in accordance with the double-logarithmic approximation. Defining
$z_h = \frac{\alem}{2\pi} \frac{1}{\eps^2} \big(\frac{\mu^2}{s}\big)^\eps$, we obtain
\begin{align}
\label{eq:F1:moments}
 F_1(\lambda) \,\simeq\,& \sum_{n=0}^\infty\; z_h^n\; h^{(n)} \;
\big\langle x^{-1-n\eps} \big\rangle_{f_c}
\Big(\frac{\mu}{m},\frac{\nu}{\sqrt{s}}\Big)\;
\big\langle y^{-1-n\eps} \big\rangle_{f_{\bar c}}
\Big(\frac{\mu}{m},\frac{\nu\sqrt{s}}{m^2}\Big)
\end{align}
in terms of the moments that we introduced in \eqref{eq:collinear:moment}, and we have made the scale dependence of these moments explicit. The moments are in general endpoint-divergent, and they generate poles in the dimensional regulator $\eps$ and the rapidity regulator $\reg$. 

We first focus on the cancellation of the rapidity divergences, which turn out to have a very specific structure. First of all, the hard function is free from any rapidity divergences, which are entirely encoded in the collinear and anti-collinear moments, as indicated by their dependence on the rapidity scale $\nu$. As argued towards the end of the previous section, these moments generate rapidity divergences for any value of $n$. These $1/\reg$ poles, together with the respective dependence on the rapidity scale $\nu$, must cancel for arbitrary values of $\eps$, i.e.~without expanding in the dimensional regulator. Interestingly, this cancellation happens not only for the sum of all contributions, but for each term in the sum individually. The reason is that the moments, which depend on powers of $(\mu/m)^\eps$, cannot compensate for the powers of $s^\eps$ in $z_h$, which forbids a cross-talk between the various contributions. The rapidity divergences must therefore cancel in the product of a collinear and an anti-collinear moment for each fixed value of $n$. One can then use the standard collinear-anomaly argument to show that the rapidity logarithms associated with the $1/\alpha$ poles exponentiate~\cite{Becher:2010tm,Becher:2011pf},
\begin{align}
\label{eq:anomaly}
\big\langle x^{-1-n\eps} \big\rangle_{f_c}
\Big(\frac{\mu}{m},\frac{\nu}{\sqrt{s}}\Big)\;
\big\langle y^{-1-n\eps} \big\rangle_{f_{\bar c}}
\Big(\frac{\mu}{m},\frac{\nu\sqrt{s}}{m^2}\Big) 
= r_n(\mu/m) \,\times\, \Big(\frac{m^2}{s}\Big)^{\mathcal{F}_n(\mu/m)}\,,
\end{align}
where $\mathcal{F}_n(\mu/m)$ is called a collinear-anomaly exponent and $r_n(\mu/m)$ a remainder function. The right-hand side of this expression is manifestly independent of the rapidity scale $\nu$, and the respective bare quantities depend only  on powers of $(\mu/m)^\eps$, as indicated by their arguments.

After inserting \eqref{eq:anomaly} into the expression for the form factor, we obtain an infinite tower of collinear-anomaly exponents, 
\begin{align}
\label{eq:F1:anomalies}
 F_1(\lambda) \,\simeq\,& \sum_{n=0}^\infty\; z_h^n\; h^{(n)} \;
r_n(\mu/m) \,\times\, 
\Big(\frac{m^2}{s}\Big)^{\mathcal{F}_n(\mu/m)}\,.
\end{align}
We next turn to the cancellation of the $1/\eps$ poles, which is more complicated, since it happens only on the level of the sum of all contributions.
To this end, we expand the collinear-anomaly exponents and the remainder functions in $\eps$, keeping again only the highest poles in the double-logarithmic approximation. The refactorization condition \eqref{eq:refactorization:DL} provides an important constraint in this context, since it tells us that $\mathcal{F}_n(\mu/m)=\mathcal{O}(\alem^{n+1})$, as argued towards the end of Sec.~\ref{subsec:refactorization}. We thus expand
\begin{align}
\label{eq:rFparam}
r_n(\mu/m) &=
\sum_{k=0}^\infty \,\left(\frac{\alem}{2\pi}\right)^k 
\Big(\frac{\mu^2}{m^2}\Big)^{k\eps} \;
\frac{r_n^{(k)}}{\eps^{2k}}\,,
\nonumber\\
\mathcal{F}_n(\mu/m) &=
\sum_{l=n+1}^\infty \,\left(\frac{\alem}{2\pi}\right)^l 
\Big(\frac{\mu^2}{m^2}\Big)^{l\eps} \;
\frac{\mathcal{F}_n^{(l)}}{\eps^{2l-1}}\,,
\end{align}
where $r_n^{(0)} =1$ at tree level, since the moments trivially evaluate to unity in the convolution with the delta function. Notice also that the power of the $\eps$-divergences is reduced by one unit for the anomaly exponents, since they are generated by an additional $1/\reg$ pole in the loop calculations.

We finally determine the unknown coefficients $h^{(n)}$, $r_n^{(k)}$ and $\mathcal{F}_n^{(l)}$ from the requirement that the form factor $F_1(\lambda)$ is finite in the limit $\eps \to 0$. Counting the free parameters and the number of consistency relations at each perturbative order shows that this requirement fixes all coefficients but one, which we choose to be $h^{(1)}$. Although the complete system of equations is non-linear, all new coefficients at $\mathcal{O}(\alem^{n})$ can be determined by solving linear relations, once all coefficients up to $\mathcal{O}(\alem^{n-1})$ are known. In this way the double-logarithmic series can be constructed order-by-order. More precisely, one has $2n$ consistency relations from pole cancellation, and $(2+3n)/2$ new parameters for even $n$, respectively $(3+3n)/2$ for odd $n$. This leaves one free parameter for $n=1$, while the number of constraints matches the number of coefficients for $n =2,3$. Interestingly, the system of linear equations is over-determined for $n\geq 4$. We then find that the coefficients in the expansion of the hard function are given by $h^{(n)} = (h^{(1)})^n/(n!)^2$, and the bare hard function thus evaluates to a modified Bessel function in the given approximation,
\begin{align}
\label{eq:hardfun:DL}
H\Big(\frac{\mu^2}{xys}\Big) &\simeq
\sum_{n=0}^\infty \,\left(\frac{\alem}{2\pi}\right)^n 
\Big(\frac{\mu^2}{x y s}\Big)^{n\eps} \;
\frac{ (h^{(1)})^n}{(n!)^2\,\eps^{2n}} 
= I_0\bigg(2\sqrt{\frac{\alem}{2\pi}\frac{h^{(1)}}{\eps^2}
\Big(\frac{\mu^2}{x y s}\Big)^{\eps}}\,\bigg)
\,.
\end{align}
We could not find closed expressions for the anomaly coefficients $\mathcal{F}_n^{(l)}$ and the ones of the remainder function $r_n^{(k)}$. For the former, we find that the lowest-order terms are given by
\begin{align}
\mathcal{F}_0^{(1)}&= - h^{(1)}, 
&  & & & & &
\\ 
\mathcal{F}_0^{(2)}&= \frac12 (h^{(1)})^2, &
\mathcal{F}_1^{(2)}&= -\frac12 (h^{(1)})^2, 
& & & &
\nonumber\\ 
\mathcal{F}_0^{(3)}&= -\frac{7}{12} (h^{(1)})^3, &
\mathcal{F}_1^{(3)}&= \frac23 (h^{(1)})^3, &
\mathcal{F}_2^{(3)}&= -\frac{1}{12} (h^{(1)})^3, 
& &
\nonumber\\ 
\mathcal{F}_0^{(4)}&= \frac{127}{144} (h^{(1)})^4, &
\mathcal{F}_1^{(4)}&= -\frac{15}{16} (h^{(1)})^4, &
\mathcal{F}_2^{(4)}&= \frac{1}{16} (h^{(1)})^4, &
\mathcal{F}_3^{(4)}&= -\frac{1}{144} (h^{(1)})^4.
\nonumber
\end{align}
The remainder-function coefficients can, however, be related to the anomaly exponents via\footnote{If the $r_n(\mu/m)$ and ${\cal F}_n(\mu/m)$ are considered as functions of the variable $z_c = \frac{\alem}{2\pi} \frac{1}{\eps^2} \big(\frac{\mu^2}{m^2}\big)^\eps$, we note that these relations amount to a linear differential equation, $$ r_n(z_c) = -\frac{1}{\eps} \frac{(n!)^2}{(h^{(1)})^{n+1} z_c^n} \, \frac{d}{dz_c} {\cal F}_n(z_c) \,. $$}
\begin{align}
r_n^{(k)} = - \frac{(n+1+k)\,(n!)^2}{(h^{(1)})^{n+1}}\,
\mathcal{F}_n^{(n+1+k)}\,,
\end{align}
which we verified explicitly up to $\mathcal{O}(\alem^{14})$.

Inserting all coefficients back into~\eqref{eq:F1:anomalies}, expanding first in the coupling $\alem$ and then in the dimensional regulator $\eps$, one  recovers order-by-order the well-known Bessel function,
\begin{align}
 F_1(\lambda) \simeq  \frac{I_1\left(2\sqrt{\frac{\alem}{2\pi} \,h^{(1)}\ln^2\lambda^2}\right)}{\sqrt{\frac{\alem}{2\pi}\,h^{(1)}\ln^2\lambda^2}} \,,
\end{align}
which still depends on an unknown coefficient $h^{(1)}$. As usual, we thus need to determine one coefficient from a one-loop calculation in order to resum the double logarithms to all orders. This coefficient is likely related to the one-loop cusp anomalous dimension, and it can be directly read off from the expression \eqref{eq:hardfun}, which yields $h^{(1)}=1$ in our conventions. We hence confirm the  result \eqref{eq:BesselI} that we obtained earlier, but it is interesting to note how the resummation is achieved within the effective theory from a non-trivial cancellation of $1/\reg$ and $1/\eps$ poles. In particular, we are not aware of any resummation in \scettwo~that proceeds via an infinite tower of collinear-anomaly exponents.

\subsection{Higher-order structure of $f_c(x\to 0)$}

The refactorization condition \eqref{eq:refactorization:DL} for the collinear function $f_c(x)$ in the limit $x\to 0$ has a similar structure as the factorization theorem \eqref{eq:fac:minimal} for the form factor $F_1(\lambda)$, and one may therefore apply similar methods to study the higher-order structure of the asymptotic function $f_c(x\to0)$ relevant in the double-logarithmic approximation. In particular, this analysis reveals that the consistency of the effective theory requires the two features that we mentioned towards the end of Sec.~\ref{subsec:refactorization}, i.e.~the bare function $f_c(x\to 0)$ generates logarithms in $\ln x$ and positive powers of $x^\eps$, starting at two-loop order. While this provides interesting insights into the endpoint dynamics at higher orders, we stress that the results from this section cannot be applied to calculate endpoint-divergent moments, since we take the limit $\reg\to 0$ in the course of this discussion.
 
The refactorization formula \eqref{eq:refactorization:DL} is indeed simply a copy of \eqref{eq:fac:minimal}, with the anti-collinear function replaced by the soft function and, consequently, the hard function replaced by the jet function.
One can thus cast the refactorization condition into a form that is similar to the starting point
\eqref{eq:DLfacscales} of the previous analysis,
\begin{align}
 f_c(x\to 0) \simeq& 
 \int_0^1 \frac{dx'}{x'}  \, f_c\Big(x';\frac{\mu}{m},\frac{\nu}{n_+p}\Big) 
\int_0^\infty \frac{d\rho}{\rho} \; 
\text{Disc}\,S\Big(\rho,x(n_+p);\frac{\mu}{m},\frac{\nu}{x(n_+p)}\Big)
J_{hc}\Big(\frac{\mu^2}{\rho x'(n_+p)}\Big),
\end{align}
where we have used the analyticity properties of the integrand in the complex $\rho$-plane to rewrite the integrations in terms of the discontinuity of the soft function -- see~\eqref{eq:disc} -- and we remind the reader of the subtlety discussed in the footnote of Sec.~\ref{subsec:refactorization}. It is also easy to understand that the functional form of the jet function agrees with the one of the hard function on the double-logarithmic level, as both functions are determined by the same massless ladder-type loop integrals. We can thus immediately exploit \eqref{eq:hardfun:DL} to derive the all-order form of the jet function in the double-logarithmic approximation,
\begin{align}
J_{hc}\Big(\frac{\mu^2}{\rho x'(n_+p)}\Big) &\simeq
I_0\bigg(2\sqrt{\frac{\alem}{2\pi}\frac{1}{\eps^2}
\Big(\frac{\mu^2}{\rho x'(n_+p)}\Big)^{\eps}}\,\bigg)
=\sum_{n=0}^\infty \,\left(\frac{\alem}{2\pi}\right)^n 
\Big(\frac{\mu^2}{\rho x'(n_+p)}\Big)^{n\eps} \;
\frac{1}{(n!)^2\,\eps^{2n}} \,.
\end{align}
Due to boost invariance, the soft function can furthermore only depend on the product $\rho x(n_+p)$, up to terms that are associated with the rapidity regulator. It is therefore convenient to introduce the dimensionless variable $\hat\rho=\rho x(n_+p)/m^2$, as well as the moments
\begin{align}
\label{eq:softmoments}
\big\langle \hat\rho^{-1-n\eps} \big\rangle_{S}\Big(\frac{\mu}{m},\frac{\nu}{x(n_+p)}\Big)
\equiv \int_0^\infty d\hat\rho \; \hat\rho^{-1-n\eps} \; 
\text{Disc}\,S\Big(\frac{m^2\hat\rho}{x(n_+p)},x(n_+p);\frac{\mu}{m},\frac{\nu}{x (n_+p)}\Big)\,.
\end{align}
We are then in the position to write down the analogous relation to \eqref{eq:F1:moments} for the collinear function in the endpoint region, 
\begin{align}
\label{eq:fc:moments}
 f_c(x\to 0) \,\simeq\,& \sum_{n=0}^\infty\; z_c^n\; \frac{x^{n\eps}}{(n!)^2} \;
\big\langle (x')^{-1-n\eps} \big\rangle_{f_c}
\Big(\frac{\mu}{m},\frac{\nu}{n_+p}\Big)\;
\big\langle \hat\rho^{-1-n\eps} \big\rangle_{S}
\Big(\frac{\mu}{m},\frac{\nu}{x(n_+p)}\Big)\,,
\end{align}
where we introduced $z_c = \frac{\alem}{2\pi} \frac{1}{\eps^2} \big(\frac{\mu^2}{m^2}\big)^\eps$. We stress that the $x$-dependence of the soft moments must be of the indicated form, since it is entirely induced by terms that violate boost invariance, and it is therefore directly tied to the rapidity scale $\nu$.

Following the discussion from the previous section, we now first consider the cancellation of the rapidity divergences, which must happen on the level of the collinear function itself, since it is a well-defined GPD in $d$ dimensions. 
Specifically, the finiteness of $f_c(x\to0)$ in the limit $\reg \to 0$ requires that the soft moments in \eqref{eq:softmoments} are endpoint-divergent; they in fact produce poles in the rapidity regulator for \emph{small} arguments $\hat{\rho}\to 0$, as discussed in App.~\ref{app:nestedintegrals}. 
By a similar argument as in the previous section, the $1/\reg$ poles must furthermore again cancel for each term in the sum individually, since the moments cannot compensate for the factors $x^{n\eps}$. We thus obtain similar to \eqref{eq:anomaly}
\begin{align}
\label{eq:fc:anomaly}
\big\langle (x')^{-1-n\eps} \big\rangle_{f_c}
\Big(\frac{\mu}{m},\frac{\nu}{n_+p}\Big)\;
\big\langle \hat\rho^{-1-n\eps} \big\rangle_{S}
\Big(\frac{\mu}{m},\frac{\nu}{x(n_+p)}\Big)
= \hat r_n(\mu/m) \,\times\, x^{\mathcal{F}_n(\mu/m)}\,,
\end{align}
where $\mathcal{F}_n(\mu/m)$ are the same anomaly exponents that we discussed in the previous section, since the relation involves the same moments of the collinear function $f_c(x)$, whereas the remainder functions $\hat r_n(\mu/m)$ are new objects. The latter can be expanded in the form
\begin{align}
\hat r_n(\mu/m) &=
\sum_{k=1}^\infty \,\left(\frac{\alem}{2\pi}\right)^k 
\Big(\frac{\mu^2}{m^2}\Big)^{k\eps} \;
\frac{\hat r_n^{(k)}}{\eps^{2k-1}}\,,
\end{align}
which differs from \eqref{eq:rFparam}, since the soft moments only start at $\mathcal{O}(\alem)$, and the new remainder functions have one divergence less in the double-logarithmic approximation, since the collinear function itself is yet to be integrated over $x$, which produces another divergence.

We finally need to determine the coefficients $\hat r_n^{(k)}$. In contrast to the form factor $F_1(\lambda)$, we cannot invoke the strict pole cancellation argument here, since the bare collinear function is \emph{not} finite in the limit $\eps\to 0$. We do know, however, that forward GPDs have a single-logarithmic evolution, and the collinear function $f_c(x)$ can therefore only receive a single pole in $\eps$ from each order in perturbation theory. This condition, together with the lowest-order coefficients $\hat r_n^{(1)}=1/(n+1)$ that need to be determined by an explicit one-loop calculation, then determines all $\hat r_n^{(k)}$. Plugging these coefficients back into \eqref{eq:fc:moments} -- without performing an expansion in $\eps$ -- then yields at one-loop order
\begin{align}
\label{eq:fcDLNLO}
    f_c^{(1)}(x\to 0) \simeq \Big(\frac{\mu^2}{m^2}\Big)^\eps \;
		\frac{1}{\eps} \,,
\end{align}
in agreement with our previous findings in \eqref{eq:1loopbarefcendpoint} in the given approximation with $\alpha\to 0$. At two loops we obtain
\begin{align}
\label{eq:fcDLNNLO}
    f_c^{(2)}(x\to 0) \simeq \Big(\frac{\mu^2}{m^2}\Big)^{2\eps} \,
		\bigg\{ \frac{x^{\eps}-1}{2 \eps^3} - \frac{\ln x}{\eps^2}\bigg\} \,,
\end{align}
which has precisely the features that we mentioned at the beginning of this section, but we can now trace the origin of the factor $x^\eps$ and the logarithm $\ln x$. While the former arises in \eqref{eq:fc:moments} from the loop expansion of the jet function, the latter is generated in \eqref{eq:fc:anomaly} through the collinear anomaly. The structure \eqref{eq:fcDLNNLO} has in fact been predicted by explicit two-loop calculations; it can be read off e.g.~from the leading terms in (8.26) of~\cite{Boer:2018mgl}, as well as from our higher-order analysis of the nested scalar integrals \eqref{eq:nestedintegrals} in App.~\ref{app:nestedintegrals}.

It is now a straightforward task to construct the higher-order terms. The three- and four-loop contributions read e.g.
\begin{align}
\label{eq:fcDLNNNLO}
    f_c^{(3)}(x\to 0) \simeq \Big(\frac{\mu^2}{m^2}\Big)^{3\eps} \,
		\bigg\{ \frac{x^{2\eps}-14x^\eps+13}{12 \eps^5} + \frac{\ln x}{\eps^4} + \frac{\ln^2x}{2\eps^3}\bigg\} \,,
\end{align}
and 
\begin{align}
\label{eq:fcDLNNNNLO}
    f_c^{(4)}(x\to 0) \simeq  \Big(\frac{\mu^2}{m^2}\Big)^{4\eps} \,
		\bigg\{ \frac{x^{3\eps}-15x^{2\eps}+339x^\eps -325}{144 \eps^7} - \frac{(3x^\eps+23)\ln x}{12 \eps^6} - \frac{3\ln^2 x}{4\eps^5} - \frac{\ln^3x}{6\eps^4}\bigg\} \,.
\end{align}
We do not aim to find a closed expression for the unexpanded  bare function $f_c(x\to0)$. However, once we expand in $\eps$ and focus on the leading UV poles, the asymptotic collinear function itself becomes a modified Bessel function, which after restoring the tree-level contribution $\delta(1-x)$ as well as the phase-space constraint $\theta(1-x)$ takes the form
\begin{align}
 f_c(x\to 0)\big\vert_{\rm UV-div} =  \delta(1-x) + \theta(1-x) \,
\frac{\alem}{2\pi\eps} \, 
\frac{I_1\left(2\sqrt{\frac{\alem}{2\pi\eps}  \ln\frac{1}{x}}\right)}{\sqrt{\frac{\alem}{2\pi\eps}  \ln\frac{1}{x}}} \,.
\label{eq:fc:UVpoles}
\end{align}
We emphasize that this representation is not suited for calculating  endpoint-divergent moments, but it is nevertheless useful, since it provides another cross-check. The expression \eqref{eq:fc:UVpoles}
can in fact be understood as the $\overline{\text{MS}}$ renormalization factor associated with the one-loop quark-to-quark splitting kernel in the small-$x$ limit, as can be verified by performing a Mellin transform,
\begin{align}
\int_0^1 dx \, x^{N-1} f_c(x\to 0)\big\vert_{\rm UV-div} = \exp\left(\frac{\alem}{2\pi\eps} \, \frac{1}{N}\right)\,, \qquad \text{for} \,\, N > 0 \,.
\end{align}
We note that the small-$x$ limit of $f_c(x)$ is determined by the singular behaviour at $N=0$ in Mellin space.

\section{Comparison to $h\to (b \bar{b})^* \to \gamma \gamma$}
\label{sec:hgg}

In many respects the problem of muon-electron backward scattering studied in this article appears very similar to the bottom-quark induced $h\to \gamma\gamma$ decay, whose factorization properties within SCET have been derived in a sequence of papers \cite{Liu:2019oav,Liu:2020tzd,Liu:2020wbn}.
In that case the form factor multiplying the leading one-loop amplitude can,  at the double-logarithmic level, be written as
\begin{align}
\label{eq:hgg2F2}
    \mathscr{F}_b(z) = {}_2F_2\Big(1,1;\frac32,2; -z/2\Big) = 2 \, \sum_{n=0}^\infty 
    \frac{n!}{(2n+2)!} \, (-2z)^n \,,
\end{align}
with $z=\frac{\alpha_s C_F}{4\pi} \, \ln^2 \frac{m_b^2}{M_h^2}$, which should be compared to the analogous approximation for the form factor $F_1(\lambda)$ in~\eqref{eq:BesselI}. 
Both processes are described in \scettwo, and rapidity divergences are caused by light but massive fermion propagators that overlap in the endpoint region between the (anti-)collinear and soft modes.
Nevertheless, the structure of the bare factorization formula in both situations turns out to be different. For the bottom-induced $h\to \gamma\gamma$ decay amplitude the bare factorization theorem takes the schematic form (the precise form can be found in eq.~(50) of~\cite{Liu:2019oav})
\begin{align}
\label{eq:Fbfac}
 H_1 \langle \gamma\gamma|O_1|h\rangle 
 + H_2 * \langle\gamma\gamma|O_2|h\rangle 
 + H_3 \, J_{hc} * S * J_{\bar hc} \,,
\end{align}
with the $*$ symbol indicating a convolution. 
For the comparison with the factorization theorem of the muon-electron scattering amplitude in~\eqref{eq:barefactheorem}, the first term in \eqref{eq:Fbfac} plays only a minor role. The second term is a convolution of a collinear matrix element -- which is related to the leading-twist photon distribution amplitude $\phi_\gamma(x)$ -- with a respective hard-matching function $H_2(x)$. This convolution is endpoint-divergent as $x \to 0$. We note that in the above equation this term arises from two identical contributions as a sum of collinear as well as anti-collinear matrix elements, which requires using a symmetric rapidity regulator. The corresponding term to compare with would be the first term $f_c * H * f_{\bar c}$ in our factorization theorem, which contains endpoint-divergent convolutions of two collinear functions in the limit $x\to 0$ and $y \to 0$. The last term in~\eqref{eq:Fbfac} is a convolution of the soft function with two hard-collinear jet functions, and should be compared with the second term $f_c * J_{hc} * S * J_{\bar hc} * f_{\bar c}$ in our bare factorization theorem.

There is a simple physical argument that explains the main difference between the two scenarios: the helicity suppression of the $h\to\gamma\gamma$ amplitude must be present in each individual term in the factorization formula. Resolving the partonic structure of the photon in terms of a collinear $(b\bar b)$ pair is a power-suppressed effect, $\langle\gamma\gamma|O_2|h\rangle \sim m_b \ll M_h$. Similarly, also the soft function must share the helicity suppression in the third term, \mbox{$S \sim m_b$} (see e.g.~(40) and (44) in~\cite{Liu:2019oav}). The power-counting thus forbids two of these functions to appear in a single term, which forces the bare factorization formula to be a \emph{sum} of endpoint-divergent moments. This is in contrast to the muon-electron scattering amplitude: neither are the (anti-)collinear, nor are the soft functions power-suppressed in $\lambda = m_\mu/\sqrt{s}$. Hence, \emph{products} of endpoint-divergent moments appear in the bare factorization theorem. This difference is reflected by several technical aspects that we briefly summarize below:

\begin{itemize}
    \item In the double-logarithmic approximation, the form factor $\mathscr{F}_b(z)$ in~\eqref{eq:hgg2F2} can be written as~\cite{Liu:2018czl}
    \begin{align}
       \mathscr{F}_b(z) &=  2 \, \int_0^1 d\xi \, \int_0^{1} d\eta \, \theta(1-\xi-\eta) \, \mathscr{F}_S(\xi\eta z) \,,
       \qquad \mbox{with $\mathscr{F}_S(\xi\eta z)=e^{-2\xi \eta z}$} \,.
       \label{hggDL}
    \end{align}
Here the function $\mathscr{F}_S(\xi\eta z)$ in the integrand arises from the standard Sudakov-type exponentiation of soft-gluon corrections to the $h\to b \bar b$ subprocess \cite{Kotsky:1997rq}. In our case, the series of radiative corrections responsible for the double logarithms rather yields \emph{nested} integrals in~\eqref{eq:nestedintegrals}, reflecting the iterative nature of the problem. The analogous formula to (\ref{hggDL}) for muon-electron backward scattering can then be written as a consistency relation 
\begin{align}
  F_1(\lambda) \simeq \mathscr{F}_1(z) = \frac{I_1(2 \sqrt z)}{\sqrt z} =
  1 + z \, \int_0^1 d\xi \, \int_0^{1}  d\eta \, \mathscr{F}_1(\xi^2 z) \, \theta(1-\xi-\eta) \, \mathscr{F}_1(\eta^2 z) \,,
  \label{eq:F1selfcons}
\end{align}
where the form factor to be calculated appears again within the integral 
on the right-hand side, with $z = \frac{\alem}{2\pi} \ln^2 \lambda^2$. To compare with~\eqref{eq:nestedintegrals} one has to identify the logarithmic variables $\xi = \ln x/\ln \lambda^2$ and $\eta=\ln y/\ln\lambda^2$, and $\theta(1-\xi-\eta)=\theta(x y - \lambda^2)$ from the discontinuity of the soft-fermion propagator.
\item In $h\to\gamma\gamma$, the additive structure of the bare factorization formula implies that only a single rapidity divergence appears in the endpoint-divergent convolutions to all orders in perturbation theory (see~(60,61) in~\cite{Liu:2019oav}).
 In contrast, the rapidity poles in muon-electron backward scattering exponentiate in inverse moments.
\item The refactorization condition for the matrix element of $O_2$ in $h\to\gamma\gamma$ is schematically given by (see (55) in \cite{Liu:2019oav}) 
\begin{align}
 [[\langle \gamma\gamma |O_2(x)|h\rangle]] \equiv \langle \gamma\gamma |O_2(x)|h\rangle\big\vert_{x\to 0} \sim J_{hc} * S \,,
\label{eq:refac:hgg}
\end{align}
whereas in our case it reads
\begin{align}
 f_c(x\to 0) \sim f_c * J_{hc} * S \,.
 \label{eq:refac:emu}
\end{align}
The refactorization properties are related to the exponentiation of rapidity poles, and reflect the iterative nature of the problem at hand, as the collinear function $f_c$ appears on both sides of this relation. The reason why an iterative structure like in~\eqref{eq:refac:emu} cannot arise in $h\to\gamma\gamma$ is again the helicity suppression.
In this sense~(\ref{eq:refac:hgg}) can be viewed as a special case of~(\ref{eq:refac:emu}). 
\item As a consequence, the $n$-loop (with $n \geq 2$) expression for the bare collinear function $f_c(x)$ at small $x$ contains explicit powers of logarithms $\ln x$ which arise from a cancellation of rapidity poles within the endpoint-divergent convolutions of the refactorization formula. These logarithms cause higher poles in the rapidity regulator, and do not appear in $h \to \gamma \gamma$ (see eqs.~(40)-(42) in \cite{Liu:2019oav} for the corresponding two-loop expressions).
\item In $h\to \gamma\gamma$ the soft function $S$ in the factorization theorem~\eqref{eq:Fbfac} vanishes if its argument goes to zero (see eq.~(45) in~\cite{Liu:2019oav}). As a consequence, the convolution integrals with the jet functions only obtain endpoint divergences from the upper integration boundary. This integration domain is removed in the renormalized factorization theorem as the integrals have a hard upper cutoff $\sim M_h$. Starting at NLO, the soft function in muon-electron backward scattering does not vanish for small arguments, see~\eqref{eq:NLOLLsoftfct}. The convolutions of the soft function with the jet functions then receive rapidity poles also from small arguments, $\rho,\omega \to 0$, i.e. endpoint singularities would not be removed by an upper cutoff.
\end{itemize}

\section{Conclusions}
\label{sec:conclusion}

In this paper we argued that the simple perturbative textbook process of muon-electron scattering in the backward direction provides an interesting laboratory to study conceptual aspects of soft-collinear factorization. In particular, the backward-scattering amplitude at large center-of-mass energies $\sqrt{s} \gg m_{e,\mu}$ is known to receive large double-logarithmic corrections of the form $\sim\alem^{n+1} \ln^{2n} m_{e,\mu}/\sqrt{s}$ that resum to a modified Bessel function \mbox{\cite{Gorshkov:1966qd,Berestetskii:1982qgu}}. In the modern SCET formulation this structure is recovered by an iterative pattern of endpoint-divergent convolution integrals.

The presence of endpoint divergences in the factorization theorem prevent the use of renormalization-group techniques to resum the logarithmic corrections to all orders. This problem arises generically in SCET at subleading power, but for the current process it shows up already at leading power. While this feature may seem surprising, we explained that it is caused by a particular soft-enhancement mechanism that is specific to the backward kinematics. The problem of endpoint divergences can therefore be studied in a particular transparent way at leading power in an abelian version of SCET. At the same time the simple QED process unveils a structure of endpoint singularities -- even at the leading double-logarithmic level -- that is more complicated than what has been discussed in the literature so far. As the soft enhancement can lift the helicity suppression, the endpoint divergences manifest themselves in a nested pattern, in contrast to other examples like the bottom-induced $h\to\gamma\gamma$ decay~\cite{Liu:2019oav,Liu:2020tzd,Liu:2020wbn}, for which the iteration  stops already at the first step because of the helicity suppression.  

By using endpoint-refactorization conditions for the collinear matrix elements and consistency relations, we showed that the double-logarithmic corrections can be resummed in the effective theory. The cancellation of the rapidity divergences generates an infinite tower of collinear-anomaly exponents that reflects the iterative structure of the endpoint divergences. While we did not attempt to derive a renormalized factorization theorem in this work, it would be interesting to understand if a rearrangement of the terms in the bare factorization formula -- in the spirit of what has been proposed for the $h \to \gamma\gamma$ decay amplitude in~\cite{Liu:2019oav,Liu:2020tzd,Liu:2020wbn} -- can also be formulated for the current process. In view of the discussion in Sec.~\ref{sec:hgg}, it seems however likely that this will require additional conceptual and technical ingredients which is beyond the scope of this work.

The main qualitative difference that distinguishes the muon-electron scattering process from other examples that have been studied in SCET recently is that the bare factorization theorem contains \emph{pro\-ducts} of endpoint-divergent moments, whereas it contains a \emph{sum} in the other cases~\cite{Liu:2019oav,Beneke:2022obx}. Although the backward kinematics seems to be very special and fine-tuned, we argue that the process mimics the general structure of hard-exclusive processes in QCD, for which products of endpoint-divergent moments that involve hadronic light-cone distribution amplitudes (LCDAs) typically arise in the factorization theorem. The present work in fact emerged from a study of $B_c \to \eta_c$ transition form factors in the limit \mbox{$m_b \gg m_c \gg \Lambda_{\rm QCD}$~\cite{Boer:2018mgl,Bell:2005gw}}. In this limit the hadronic states can be approximated by non-relativistic bound states of two massive quarks that are on-shell and move with the same four-velocity. The charm-quark mass $m_c$ provides an intrinsic IR regulator, and relativistic corrections to the light-cone distribution amplitudes can be computed perturbatively~\cite{Bell:2008er}. Endpoint divergences then arise in inverse moments of the $B_c$-meson and $\eta_c$-meson LCDAs, and their perturbative evaluation requires a rapidity regulator. Although the physics seems to be quite different, the $B$-meson decay in this setup and the muon-electron backward-scattering amplitude share many similarities from the factorization point of view, like the exponentiation of rapidity logarithms in endpoint-divergent moments of the heavy- and light-meson LCDAs~\cite{Boer:2018mgl}. While the transition from the perturbative calculation with massive quarks to realistic charmless $B\to\pi$ form factors is non-trivial, we expect that the structure of the renormalized factorization theorem -- which is yet to be determined -- will be identical, with the soft and collinear matrix elements becoming non-perturbative objects.

In summary, we believe that our work can give new insights to the problem of endpoint-divergent convolution integrals in SCET. The considered QED process of muon-electron backward scattering is, on the one hand, simple enough to clearly illustrate the key problem without additional complications like e.g.~operator mixing, while it is, on the other hand, sufficiently generic and mimics the factorization of hard-exclusive processes in QCD.

\subsubsection*{Acknowledgements}
We thank Martin Beneke and Matthias Neubert for discussions. We are particularly thankful to Alexei Pivovarov for pointing us to references~\cite{Gorshkov:1966qd,Berestetskii:1982qgu}.
The research of G.B.\ and T.F.\ was supported by the Deutsche Forschungsgemeinschaft (DFG, German Research Foundation) under grant 396021762 - TRR 257, the research of P.B.\ by the Cluster of Excellence PRISMA$^+$ funded by DFG within the German Excellence Strategy (Project ID 39083149). G.B.\ thanks the Department of Fundamental Physics at the University of Salamanca for hospitality during the final stage of this project.

\appendix

\section{One-loop IR divergences}
\label{app:imagpart}

In~\eqref{eq:F1oneloop} we found that the NLO corrections to the muon-electron scattering amplitude generate an IR divergence,
\begin{align}
\label{eq:F1IRdiv}
    F_1^{(1)}(\lambda) = \frac{2\pi i}{\eps_{\rm IR}} + \ldots \,,
\end{align}
which must be understood as a soft singularity, since collinear singularities are regularized by the lepton masses. The reason why this IR singularity is purely imaginary is subtle. In this appendix we briefly explain how it arises from soft-photon exchanges with momenta $k^\mu\sim E_\gamma\ll m$, where $E_\gamma$ is the experimental resolution energy.

The soft physics associated with the scale $E_\gamma$ is described by the vacuum matrix element
\begin{align}
\label{eq:softME}
 \hat{S} = \bra{0} (\bar{S}_{n_-}S^\dagger_{n_-} \, \bar{S}_{n_+} S^\dagger_{n_+})(0) \ket{0} \,,
\end{align}
with light-like soft Wilson lines associated with the incoming leptons
\begin{align}
    \bar{S}_n(x) = \exp \left(+i e Q_\ell \int_{-\infty}^0 ds \, n\cdot A_s(x+sn) \right) \,,
\end{align}
and the outgoing leptons
\begin{align}
        S^\dagger_n(x) = \exp \left(+i e Q_\ell \int_0^{\infty} ds \, n\cdot A_s(x+sn) \right) \,,
\end{align}
respectively. Here we use the convention $iD^\mu = i \partial^\mu + e Q_\ell A^\mu$ for the covariant derivative in QED, with $e = \sqrt{4\pi\alem}$ and the electric charge $Q_\ell = -1$ for leptons. The perturbative calculation of the quantity $\hat{S}$ in~\eqref{eq:softME} yields scaleless integrals to all orders in dimensional regularization, and the soft function $\hat{S}$ has therefore been dropped in the discussion in the main text. Nevertheless, it is important that the effective theory correctly reproduces the IR physics of the scattering amplitude or, equivalently, that the remaining IR divergences of the hard function $H(xys)$ cancel against the UV divergences of the soft function $\hat{S}$. 

The important observation is that the product $(\bar{S} S^\dagger)$ describes a light-like Wilson line that extends from negative to positive infinity. It does not combine to unity, because the eikonal propagators associated to $\bar{S}$ and $S^\dagger$ have different $i0$-prescriptions, which is at the origin of the imaginary part. To see this, we extract the UV singularities of $\hat{S}$ at $\mathcal{O}(\alem)$, using an off-shell regularization scheme in the spirit of~\cite{Beneke:2019slt,Beneke:2020vnb}. We then find for the one-loop expression,
\begin{align}
    \hat{S}^{(1)} &= -16 i \pi^2 \int \frac{d^dk}{(2\pi)^d} \frac{1}{k^2+i0} \left( \frac{1}{n_-k - \delta} - \frac{1}{n_-k+\delta} \right) \left( \frac{1}{n_+k-\bar{\delta}}- \frac{1}{n_+k + \bar{\delta}}\right) \nonumber \\
    &= -\frac{2\pi i}{\eps_{\rm UV}} + \mathcal{O}(\eps^0)\,.
\end{align}
Here $\delta = p^2/(n_+p) + i0$ and $\bar{\delta} = \bar{p}^2/(n_-\bar{p}) +i0$ are remnants of the off-shell regularization in the soft Wilson lines. The UV poles of $\hat{S}$ thus precisely cancel against the IR poles in~\eqref{eq:F1IRdiv} as expected.

\section{Method-of-regions analysis of nested integrals}
\label{app:nestedintegrals}

We saw in Sec.~\ref{sec:resummation} that the factorization theorem and the refactorization conditions not only determine the double logarithms of the scattering amplitude to all orders by consistency, but they also fix the hard and collinear functions in this approximation, see \eqref{eq:hardfun:DL} and \eqref{eq:fcDLNLO}~-~\eqref{eq:fcDLNNNNLO}. In this appendix we confirm some of these higher-order results independently, by starting from the nested-integral representation of the double-logarithmic amplitude in~\eqref{eq:nestedintegrals}. As this set of scalar integrals originates from the relevant Feynman integrals, they obey the same type of factorization formula as the form factor $F_1(\lambda)$ in~\eqref{eq:fac:minimal}, at least on the double-logarithmic level. This allows us to extract the individual functions entering the factorization formula to a high order in $\alem$ by means of a method-of-regions analysis. To this end, we first generalize the expression \eqref{eq:nestedintegrals} to $d = 4-2\eps$ dimensions,
\begin{align}
\label{eq:nestedintegralseps}
 F_1^{(n)}(\lambda) \simeq &\lim_{\eps \to 0} \left(\frac{\mu^2}{s}\right)^{n\eps} \int \frac{dx_1}{x_1^{1+\eps}} \int \frac{dy_1}{y_1^{1+\eps}} \,\dots\, \int \frac{dx_n}{x_n^{1+\eps}} \int \frac{dy_n}{y_n^{1+\eps}} \; 
\theta(x_1 y_1-\lambda^2)\,\dots\, \theta(x_n y_n -\lambda^2) \nonumber \\
	 \times\, & 
	\theta(1-y_1) \, \theta(y_1-y_2) \, \dots \, \theta(y_n-\lambda^2) \; 
	\theta(1-x_n) \, \theta(x_n-x_{n-1}) \, \dots \, \theta(x_1-\lambda^2)  \,.
\end{align}
Here an additional factor $(x_i y_i - \lambda^2)^{-\eps}$ comes from integrating the perpendicular momentum components in each loop integral in $d$ dimensions, see e.g.~\eqref{eq:DLF1NLO} for the corresponding one-loop expression. On the double-logarithmic level, this factor can be approximated by $(x_i y_i - \lambda^2)^{-\eps} \approx (x_i y_i)^{-\eps}$, since (i) in the hard region the power counting implies $x_i y_i \gg \lambda^2$, and (ii) in the soft and collinear regions we are only interested in the UV singularities that arise in the limit $x_i y_i \to \infty$. We also simplified the prefactor in \eqref{eq:nestedintegralseps} for convenience.

A proper method-of-regions analysis of the integrals in~\eqref{eq:nestedintegralseps} requires analytic regularization. In this appendix, however, we are rather interested in the higher-order structure of the individual functions, ignoring for the moment that they enter endpoint-divergent convolutions. It is furthermore worth emphasizing that it is not the propagators that are expanded in the individual momentum regions, but rather the phase-space constraints.

\paragraph{Hard and jet functions:}
Expanding~\eqref{eq:nestedintegralseps} in the hard region with $x_i \sim y_i \sim \mathcal{O}(1) \gg \lambda^2$, we can easily derive a closed analytic expression for the highest poles in $\eps$ of the bare hard function to all orders in $\alem$. The $n$-loop contribution reads
\begin{align}
 H^{(n)}(xys) &\simeq \left( \frac{\mu^2}{s}\right)^{n\eps} \int_0^x \frac{dx_n}{x_n^{1+\eps}} \, \dots 
  \int_0^{x_2} \frac{dx_1}{x_1^{1+\eps}}
  \times \int_0^y \frac{dy_1}{y_1^{1+\eps}} \, \dots
  \int_0^{y_{n-1}} \frac{dy_n}{y_n^{1+\eps}} \nonumber \\
  &= \left( \frac{\mu^2}{xys}\right)^{n\eps} \frac{1}{n!^2 \eps^{2n}} \,,
\end{align}
which sums to a modified Bessel function and confirms the result in~\eqref{eq:hardfun:DL}. The hard-collinear jet function has precisely the same functional form in this approximation, but with different arguments.

\paragraph{Collinear functions:}
As argued before, we are only interested here in the expressions for the collinear functions with the rapidity regulator set to zero, which can be compared to \eqref{eq:fcDLNLO}~-~\eqref{eq:fcDLNNNNLO}. Hence, the collinear and anti-collinear functions can be identified upon replacing the argument $x \to y$.  Adopting the power counting in the collinear region, $x_i \sim \mathcal{O}(1)$ and $y_i \sim \mathcal{O}(\lambda^2)$, and dropping the integrations that correspond to the convolution with the tree-level hard and anti-collinear functions, we can write the $n$-loop contribution to the collinear function in the given approximation as
\begin{align}
\label{eq:fcDLnloop}
 f^{(n)}_c(x) \simeq \left( \frac{\mu^2}{s}\right)^{n\eps} &\int_0^1 \frac{dx_1}{x_1^{\eps}} \delta(x_1-x) \int^{1}_{x_1} \frac{dx_2}{x_2^{1+\eps}} \, \dots  \int^{1}_{x_{n-1}} \frac{dx_n}{x_n^{1+\eps}} \nonumber \\
  \times &\int^\infty_{\lambda^2/x_1} \frac{dy_1}{y_1^{1+\eps}} \int^{y_1}_{\lambda^2/x_2} \frac{dy_2}{y_2^{1+\eps}} \, \dots \int^{y_{n-1}}_{\lambda^2/x_n} \frac{dy_n}{y_n^{1+\eps}} \,.
\end{align}
We do not aim to find a closed form for the $n$-loop integral, but we rather evaluate this expression order-by-order in $\alem$. At one-loop order this gives
\begin{align}
 f^{(1)}_c(x) &\simeq \left( \frac{\mu^2}{s}\right)^{\eps} \int_0^1 \frac{dx_1}{x_1^{\eps}} \delta(x_1-x)
 \int^\infty_{\lambda^2/x_1} \frac{dy_1}{y_1^{1+\eps}} = \left( \frac{\mu^2}{m^2}\right)^{\eps} \frac{1}{\eps} \,,
\end{align}
in agreement with~\eqref{eq:fcDLNLO}, if we again omit the step functions that constrain the integration domain to the unit interval. Starting at two loops, the structures become more interesting,
\begin{align}
 f^{(2)}_c(x) &\simeq \left( \frac{\mu^2}{s}\right)^{2\eps} \int_0^1 \frac{dx_1}{x_1^{\eps}} \delta(x_1-x) \int^{1}_{x_1} \frac{dx_2}{x_2^{1+\eps}}
 \int^\infty_{\lambda^2/x_1} \frac{dy_1}{y_1^{1+\eps}} \int^{y_1}_{\lambda^2/x_2} \frac{dy_2}{y_2^{1+\eps}} \nonumber \\
 &= \left( \frac{\mu^2}{m^2}\right)^{2\eps} \left(-\frac{1}{2\eps^2}\right) \, \int^{1}_{x} \frac{dx_2}{x_2} \bigg\{\left(\frac{x}{x_2} \right)^\eps-2\bigg\} \nonumber \\
 &= \left( \frac{\mu^2}{m^2}\right)^{2\eps} \bigg\{\frac{x^\eps-1}{2\eps^3} - \frac{\ln x}{\eps^2} \bigg\} \,,
\end{align}
which confirms~\eqref{eq:fcDLNNLO}. In particular, this analysis shows that the explicit logarithm $\ln x$ arises because $x$ enters as a cutoff of intermediate longitudinal momentum integrals, and it does \emph{not} arise from an expansion in the dimensional regulator $\eps$. It is a straightforward task to verify along these lines also the three- and four-loop results in~\eqref{eq:fcDLNNNLO} and~\eqref{eq:fcDLNNNNLO} that were obtained in the main text from consistency relations. 

\paragraph{Soft function:} Lastly, we adopt the soft counting $x_i \sim y_i \sim \mathcal{O}(\lambda)$, which leads to the following $n$-loop representation of the soft function in the double-logarithmic approximation,
\begin{align}
    \text{Disc} \, S^{(n)}(\rho\omega) \simeq \left(\frac{\mu^2}{s}\right)^{n\eps} &\int_0^\infty \frac{dx_1}{x_1^\eps} \, \delta(x_1-\omega/\sqrt{s}) \int_0^\infty \frac{dy_n}{y_n^\eps} \, \delta(y_n-\rho/\sqrt{s}) \nonumber \\
    \times \,&\int \frac{dx_2}{x_2^{1+\eps}} \dots \frac{dx_n}{x_n^{1+\eps}} \int \frac{dy_1}{y_1^{1+\eps}} \dots \frac{dy_{n-1}}{y_{n-1}^{1+\eps}} \theta(x_1 y_1 - \lambda^2) \dots \theta(x_n y_n - \lambda^2) \nonumber \\
    \times \; &\theta(x_n - x_{n-1}) \dots \theta(x_2-x_1) \times \theta(y_1-y_2) \dots \theta(y_{n-1}-y_n) \,.
\end{align}
We remark that the nested integrals already involve the discontinuity of the massive lepton propagators, and hence the soft region corresponds to the discontinuity of the soft function as well, $\text{Disc} \, S(\rho\omega) =  S(\rho\omega+i0) - S(\rho\omega-i0)$, see the discussion around~\eqref{eq:disc}.
We now compute the first two terms in the perturbative series explicitly. At LO we have
\begin{align}
    \text{Disc} \, S^{(1)}(\rho\omega) &\simeq \left(\frac{\mu^2}{s}\right)^{\eps} \int_0^\infty \frac{dx_1}{x_1^\eps} \, \delta(x_1-\omega/\sqrt{s}) \int_0^\infty \frac{dy_1}{y_1^\eps} \, \delta(y_1-\rho/\sqrt{s}) \; \theta(x_1y_1-\lambda^2) \nonumber \\
    &= \left(\frac{\mu^2}{\rho\omega}\right)^\eps \theta(\rho\omega-m^2) \,.
\end{align}
This expression is in agreement with the discontinuity of the result in~\eqref{eq:soft:treelevel} in the double-logarithmic approximation and for vanishing rapidity regulator $\alpha \to 0$, since the singularities from the convolution integrals arise from the limit $\rho\omega \to \infty$. At NLO we get
\begin{align}
\label{eq:NLOLLsoftfct}
    \text{Disc} \, S^{(2)}(\rho\omega) \simeq &\left(\frac{\mu^2}{s}\right)^{2\eps} \int_0^\infty \frac{dx_1}{x_1^\eps} \, \delta(x_1-\omega/\sqrt{s}) \int_0^\infty \frac{dy_2}{y_2^\eps} \, \delta(y_2-\rho/\sqrt{s}) \int \frac{dx_2}{x_2^{1+\eps}} \int \frac{dy_1}{y_1^{1+\eps}} \nonumber \\
    &\,\times\; \theta(x_1 y_1 - \lambda^2) \;\theta(x_2 y_2 - \lambda^2) \; \theta(x_2-x_1) \;\theta(y_1-y_2) \nonumber \\
    = &\left(\frac{\mu^2}{\rho\omega}\right)^{2\eps}
    \left(\int \frac{du}{u^{1+\eps}} \,\theta\left(u - \frac{m^2}{\rho\omega}\right) \, \theta(u-1)\right)^2 
    \nonumber \\
    = \, &\frac{1}{\eps^2} \left(\frac{\mu^2}{\rho\omega}\right)^{2\eps} \theta(\rho\omega-m^2) + \frac{1}{\eps^2} \left(\frac{\mu^2}{m^2}\right)^{2\eps} \theta(m^2-\rho\omega) \,.
\end{align}
This result shows that the soft function has in general two branches, one for $\rho\omega>m^2$ and one for $\rho\omega<m^2$. In particular, the latter does \emph{not} vanish in the limit $\rho\omega\to 0$, which implies that the convolution of the soft function with the hard-collinear jet functions is again endpoint-divergent (and requires rapidity regulators) for \emph{small} arguments. As we work in \scettwo \,, this does not mean that a new region with even lower virtuality appears. Instead, it precisely reflects the kinematic configuration associated with the leading double logarithms described in Sec.~\ref{subsec:DLamplitude}: all fermion propagators go on-shell with virtuality $k^2 \sim m^2$, and are strongly ordered in their rapidity, such that all photon propagators become eikonal and would be identified as matching coefficients in a refactorization formula. We therefore expect that the soft function for small arguments obeys again a refactorization formula, and at higher loops receives logarithmic corrections of the form $\ln(\rho\omega/m^2)$.

Lastly, we mention that the soft function for asymptotically large arguments $S_\infty(\rho\omega) \equiv \text{Disc} \, S(\rho \omega \to \infty)$, which plays an important part in the rearrangement of the factorization formula in~\cite{Liu:2019oav}, can be given in a closed form
\begin{align}
\label{eq:Sinfty}
    S_{\infty}(\rho\omega) \simeq \frac{\alem}{2\pi} \left(\frac{\mu^2}{\rho\omega}\right)^\eps I_0\left( 2\sqrt{\frac{\alem}{2\pi} \frac{1}{\eps^2} \left(\frac{\mu^2}{\rho\omega}\right)^\eps}\right) \,,
\end{align}
which again involves a modified Bessel function.

\bibliography{refs}

\providecommand{\href}[2]{#2}\begingroup\raggedright\begin{thebibliography}{10}

\bibitem{Bauer:2000yr}
C.~W. Bauer, S.~Fleming, D.~Pirjol and I.~W. Stewart, \emph{{An Effective field
  theory for collinear and soft gluons: Heavy to light decays}},
  \href{http://dx.doi.org/10.1103/PhysRevD.63.114020}{\emph{Phys. Rev. D} {\bf
  63} (2001) 114020}, [\href{http://arxiv.org/abs/hep-ph/0011336}{{\tt
  hep-ph/0011336}}].

\bibitem{Bauer:2001yt}
C.~W. Bauer, D.~Pirjol and I.~W. Stewart, \emph{{Soft collinear factorization
  in effective field theory}},
  \href{http://dx.doi.org/10.1103/PhysRevD.65.054022}{\emph{Phys. Rev. D} {\bf
  65} (2002) 054022}, [\href{http://arxiv.org/abs/hep-ph/0109045}{{\tt
  hep-ph/0109045}}].

\bibitem{Beneke:2002ph}
M.~Beneke, A.~P. Chapovsky, M.~Diehl and T.~Feldmann, \emph{{Soft collinear
  effective theory and heavy to light currents beyond leading power}},
  \href{http://dx.doi.org/10.1016/S0550-3213(02)00687-9}{\emph{Nucl. Phys. B}
  {\bf 643} (2002) 431--476}, [\href{http://arxiv.org/abs/hep-ph/0206152}{{\tt
  hep-ph/0206152}}].

\bibitem{Beneke:2002ni}
M.~Beneke and T.~Feldmann, \emph{{Multipole expanded soft collinear effective
  theory with non-Abelian gauge symmetry}},
  \href{http://dx.doi.org/10.1016/S0370-2693(02)03204-5}{\emph{Phys. Lett. B}
  {\bf 553} (2003) 267--276}, [\href{http://arxiv.org/abs/hep-ph/0211358}{{\tt
  hep-ph/0211358}}].

\bibitem{Larkoski:2014bxa}
A.~J. Larkoski, D.~Neill and I.~W. Stewart, \emph{{Soft Theorems from Effective
  Field Theory}}, \href{http://dx.doi.org/10.1007/JHEP06(2015)077}{\emph{JHEP}
  {\bf 06} (2015) 077}, [\href{http://arxiv.org/abs/1412.3108}{{\tt
  1412.3108}}].

\bibitem{Moult:2017rpl}
I.~Moult, I.~W. Stewart and G.~Vita, \emph{{A subleading operator basis and
  matching for $gg \to H$}},
  \href{http://dx.doi.org/10.1007/JHEP07(2017)067}{\emph{JHEP} {\bf 07} (2017)
  067}, [\href{http://arxiv.org/abs/1703.03408}{{\tt 1703.03408}}].

\bibitem{Feige:2017zci}
I.~Feige, D.~W. Kolodrubetz, I.~Moult and I.~W. Stewart, \emph{{A Complete
  Basis of Helicity Operators for Subleading Factorization}},
  \href{http://dx.doi.org/10.1007/JHEP11(2017)142}{\emph{JHEP} {\bf 11} (2017)
  142}, [\href{http://arxiv.org/abs/1703.03411}{{\tt 1703.03411}}].

\bibitem{Chang:2017atu}
C.-H. Chang, I.~W. Stewart and G.~Vita, \emph{{A Subleading Power Operator
  Basis for the Scalar Quark Current}},
  \href{http://dx.doi.org/10.1007/JHEP04(2018)041}{\emph{JHEP} {\bf 04} (2018)
  041}, [\href{http://arxiv.org/abs/1712.04343}{{\tt 1712.04343}}].

\bibitem{Beneke:2017ztn}
M.~Beneke, M.~Garny, R.~Szafron and J.~Wang, \emph{{Anomalous dimension of
  subleading-power N-jet operators}},
  \href{http://dx.doi.org/10.1007/JHEP03(2018)001}{\emph{JHEP} {\bf 03} (2018)
  001}, [\href{http://arxiv.org/abs/1712.04416}{{\tt 1712.04416}}].

\bibitem{Moult:2018jjd}
I.~Moult, I.~W. Stewart, G.~Vita and H.~X. Zhu, \emph{{First Subleading Power
  Resummation for Event Shapes}},
  \href{http://dx.doi.org/10.1007/JHEP08(2018)013}{\emph{JHEP} {\bf 08} (2018)
  013}, [\href{http://arxiv.org/abs/1804.04665}{{\tt 1804.04665}}].

\bibitem{Beneke:2018rbh}
M.~Beneke, M.~Garny, R.~Szafron and J.~Wang, \emph{{Anomalous dimension of
  subleading-power $N$-jet operators. Part II}},
  \href{http://dx.doi.org/10.1007/JHEP11(2018)112}{\emph{JHEP} {\bf 11} (2018)
  112}, [\href{http://arxiv.org/abs/1808.04742}{{\tt 1808.04742}}].

\bibitem{Beneke:2018gvs}
M.~Beneke, A.~Broggio, M.~Garny, S.~Jaskiewicz, R.~Szafron, L.~Vernazza et~al.,
  \emph{{Leading-logarithmic threshold resummation of the Drell-Yan process at
  next-to-leading power}},
  \href{http://dx.doi.org/10.1007/JHEP03(2019)043}{\emph{JHEP} {\bf 03} (2019)
  043}, [\href{http://arxiv.org/abs/1809.10631}{{\tt 1809.10631}}].

\bibitem{Ebert:2018gsn}
M.~A. Ebert, I.~Moult, I.~W. Stewart, F.~J. Tackmann, G.~Vita and H.~X. Zhu,
  \emph{{Subleading power rapidity divergences and power corrections for
  q$_{T}$}}, \href{http://dx.doi.org/10.1007/JHEP04(2019)123}{\emph{JHEP} {\bf
  04} (2019) 123}, [\href{http://arxiv.org/abs/1812.08189}{{\tt 1812.08189}}].

\bibitem{Beneke:2019kgv}
M.~Beneke, M.~Garny, R.~Szafron and J.~Wang, \emph{{Violation of the
  Kluberg-Stern-Zuber theorem in SCET}},
  \href{http://dx.doi.org/10.1007/JHEP09(2019)101}{\emph{JHEP} {\bf 09} (2019)
  101}, [\href{http://arxiv.org/abs/1907.05463}{{\tt 1907.05463}}].

\bibitem{Beneke:2019mua}
M.~Beneke, M.~Garny, S.~Jaskiewicz, R.~Szafron, L.~Vernazza and J.~Wang,
  \emph{{Leading-logarithmic threshold resummation of Higgs production in gluon
  fusion at next-to-leading power}},
  \href{http://dx.doi.org/10.1007/JHEP01(2020)094}{\emph{JHEP} {\bf 01} (2020)
  094}, [\href{http://arxiv.org/abs/1910.12685}{{\tt 1910.12685}}].

\bibitem{Moult:2019uhz}
I.~Moult, I.~W. Stewart, G.~Vita and H.~X. Zhu, \emph{{The Soft Quark
  Sudakov}}, \href{http://dx.doi.org/10.1007/JHEP05(2020)089}{\emph{JHEP} {\bf
  05} (2020) 089}, [\href{http://arxiv.org/abs/1910.14038}{{\tt 1910.14038}}].

\bibitem{Liu:2019oav}
Z.~L. Liu and M.~Neubert, \emph{{Factorization at subleading power and
  endpoint-divergent convolutions in $h\to\gamma\gamma$ decay}},
  \href{http://dx.doi.org/10.1007/JHEP04(2020)033}{\emph{JHEP} {\bf 04} (2020)
  033}, [\href{http://arxiv.org/abs/1912.08818}{{\tt 1912.08818}}].

\bibitem{Wang:2019mym}
J.~Wang, \emph{{Resummation of double logarithms in loop-induced processes with
  effective field theory}},  \href{http://arxiv.org/abs/1912.09920}{{\tt
  1912.09920}}.

\bibitem{Beneke:2020ibj}
M.~Beneke, M.~Garny, S.~Jaskiewicz, R.~Szafron, L.~Vernazza and J.~Wang,
  \emph{{Large-x resummation of off-diagonal deep-inelastic parton scattering
  from d-dimensional refactorization}},
  \href{http://dx.doi.org/10.1007/JHEP10(2020)196}{\emph{JHEP} {\bf 10} (2020)
  196}, [\href{http://arxiv.org/abs/2008.04943}{{\tt 2008.04943}}].

\bibitem{Liu:2020tzd}
Z.~L. Liu, B.~Mecaj, M.~Neubert and X.~Wang, \emph{{Factorization at subleading
  power, Sudakov resummation, and endpoint divergences in soft-collinear
  effective theory}},
  \href{http://dx.doi.org/10.1103/PhysRevD.104.014004}{\emph{Phys. Rev. D} {\bf
  104} (2021) 014004}, [\href{http://arxiv.org/abs/2009.04456}{{\tt
  2009.04456}}].

\bibitem{Liu:2020wbn}
Z.~L. Liu, B.~Mecaj, M.~Neubert and X.~Wang, \emph{{Factorization at subleading
  power and endpoint divergences in $h\to\gamma\gamma$ decay. Part II.
  Renormalization and scale evolution}},
  \href{http://dx.doi.org/10.1007/JHEP01(2021)077}{\emph{JHEP} {\bf 01} (2021)
  077}, [\href{http://arxiv.org/abs/2009.06779}{{\tt 2009.06779}}].

\bibitem{Beneke:2022obx}
M.~Beneke, M.~Garny, S.~Jaskiewicz, J.~Strohm, R.~Szafron, L.~Vernazza et~al.,
  \emph{{Next-to-leading power endpoint factorization and resummation for
  off-diagonal ''gluon'' thrust}},  \href{http://arxiv.org/abs/2205.04479}{{\tt
  2205.04479}}.

\bibitem{Chay:2002vy}
J.~Chay and C.~Kim, \emph{{Collinear effective theory at subleading order and
  its application to heavy - light currents}},
  \href{http://dx.doi.org/10.1103/PhysRevD.65.114016}{\emph{Phys. Rev. D} {\bf
  65} (2002) 114016}, [\href{http://arxiv.org/abs/hep-ph/0201197}{{\tt
  hep-ph/0201197}}].

\bibitem{Manohar:2002fd}
A.~V. Manohar, T.~Mehen, D.~Pirjol and I.~W. Stewart, \emph{{Reparameterization
  invariance for collinear operators}},
  \href{http://dx.doi.org/10.1016/S0370-2693(02)02029-4}{\emph{Phys. Lett. B}
  {\bf 539} (2002) 59--66}, [\href{http://arxiv.org/abs/hep-ph/0204229}{{\tt
  hep-ph/0204229}}].

\bibitem{Pirjol:2002km}
D.~Pirjol and I.~W. Stewart, \emph{{A Complete basis for power suppressed
  collinear ultrasoft operators}},
  \href{http://dx.doi.org/10.1103/PhysRevD.69.019903}{\emph{Phys. Rev. D} {\bf
  67} (2003) 094005}, [\href{http://arxiv.org/abs/hep-ph/0211251}{{\tt
  hep-ph/0211251}}].

\bibitem{Boer:2018mgl}
P.~B\"oer, \emph{{QCD Factorisation in Exclusive Semileptonic B Decays: New
  Applications and Resummation of Rapidity Logarithms}}.
\newblock PhD thesis, University of Siegen, 2018.
\newblock
  \href{https://dspace.ub.uni-siegen.de/handle/ubsi/1369}{https://dspace.ub.uni-siegen.de/handle/ubsi/1369}.

\bibitem{Gorshkov:1966qd}
V.~G. Gorshkov, V.~N. Gribov, L.~N. Lipatov and G.~V. Frolov, \emph{{Double
  logarithmic asymptotics of quantum electrodynamics}},
  \href{http://dx.doi.org/10.1016/0031-9163(66)90701-3}{\emph{Phys. Lett.} {\bf
  22} (1966) 671--673}.

\bibitem{Berestetskii:1982qgu}
V.~B. Berestetskii, E.~M. Lifshitz and L.~P. Pitaevskii, \emph{{Quantum
  Electrodynamics}}, vol.~4 of \emph{Course of Theoretical Physics}.
\newblock Pergamon Press, Oxford, 1982.

\bibitem{Bell:2005gw}
G.~Bell and T.~Feldmann, \emph{{Heavy-to-light form-factors for
  non-relativistic bound states}},
  \href{http://dx.doi.org/10.1016/j.nuclphysbps.2006.11.053}{\emph{Nucl. Phys.
  B Proc. Suppl.} {\bf 164} (2007) 189--192},
  [\href{http://arxiv.org/abs/hep-ph/0509347}{{\tt hep-ph/0509347}}].

\bibitem{Bern:2000ie}
Z.~Bern, L.~J. Dixon and A.~Ghinculov, \emph{{Two loop correction to Bhabha
  scattering}}, \href{http://dx.doi.org/10.1103/PhysRevD.63.053007}{\emph{Phys.
  Rev. D} {\bf 63} (2001) 053007},
  [\href{http://arxiv.org/abs/hep-ph/0010075}{{\tt hep-ph/0010075}}].

\bibitem{Bonciani:2021okt}
R.~Bonciani et~al., \emph{{Two-Loop Four-Fermion Scattering Amplitude in QED}},
  \href{http://dx.doi.org/10.1103/PhysRevLett.128.022002}{\emph{Phys. Rev.
  Lett.} {\bf 128} (2022) 022002}, [\href{http://arxiv.org/abs/2106.13179}{{\tt
  2106.13179}}].

\bibitem{CarloniCalame:2020yoz}
C.~M. Carloni~Calame, M.~Chiesa, S.~M. Hasan, G.~Montagna, O.~Nicrosini and
  F.~Piccinini, \emph{{Towards muon-electron scattering at NNLO}},
  \href{http://dx.doi.org/10.1007/JHEP11(2020)028}{\emph{JHEP} {\bf 11} (2020)
  028}, [\href{http://arxiv.org/abs/2007.01586}{{\tt 2007.01586}}].

\bibitem{Banerjee:2020rww}
P.~Banerjee, T.~Engel, A.~Signer and Y.~Ulrich, \emph{{QED at NNLO with
  McMule}}, \href{http://dx.doi.org/10.21468/SciPostPhys.9.2.027}{\emph{SciPost
  Phys.} {\bf 9} (2020) 027}, [\href{http://arxiv.org/abs/2007.01654}{{\tt
  2007.01654}}].

\bibitem{Becher:2011dz}
T.~Becher and G.~Bell, \emph{{Analytic Regularization in Soft-Collinear
  Effective Theory}},
  \href{http://dx.doi.org/10.1016/j.physletb.2012.05.016}{\emph{Phys. Lett. B}
  {\bf 713} (2012) 41--46}, [\href{http://arxiv.org/abs/1112.3907}{{\tt
  1112.3907}}].

\bibitem{Chiu:2009yx}
J.-y. Chiu, A.~Fuhrer, A.~H. Hoang, R.~Kelley and A.~V. Manohar,
  \emph{{Soft-Collinear Factorization and Zero-Bin Subtractions}},
  \href{http://dx.doi.org/10.1103/PhysRevD.79.053007}{\emph{Phys. Rev. D} {\bf
  79} (2009) 053007}, [\href{http://arxiv.org/abs/0901.1332}{{\tt 0901.1332}}].

\bibitem{Chiu:2012ir}
J.-Y. Chiu, A.~Jain, D.~Neill and I.~Z. Rothstein, \emph{{A Formalism for the
  Systematic Treatment of Rapidity Logarithms in Quantum Field Theory}},
  \href{http://dx.doi.org/10.1007/JHEP05(2012)084}{\emph{JHEP} {\bf 05} (2012)
  084}, [\href{http://arxiv.org/abs/1202.0814}{{\tt 1202.0814}}].

\bibitem{Echevarria:2015byo}
M.~G. Echevarria, I.~Scimemi and A.~Vladimirov, \emph{{Universal transverse
  momentum dependent soft function at NNLO}},
  \href{http://dx.doi.org/10.1103/PhysRevD.93.054004}{\emph{Phys. Rev. D} {\bf
  93} (2016) 054004}, [\href{http://arxiv.org/abs/1511.05590}{{\tt
  1511.05590}}].

\bibitem{Li:2016axz}
Y.~Li, D.~Neill and H.~X. Zhu, \emph{{An exponential regulator for rapidity
  divergences}},
  \href{http://dx.doi.org/10.1016/j.nuclphysb.2020.115193}{\emph{Nucl. Phys. B}
  {\bf 960} (2020) 115193}, [\href{http://arxiv.org/abs/1604.00392}{{\tt
  1604.00392}}].

\bibitem{Radyushkin:1997ki}
A.~V. Radyushkin, \emph{{Nonforward parton distributions}},
  \href{http://dx.doi.org/10.1103/PhysRevD.56.5524}{\emph{Phys. Rev. D} {\bf
  56} (1997) 5524--5557}, [\href{http://arxiv.org/abs/hep-ph/9704207}{{\tt
  hep-ph/9704207}}].

\bibitem{Ji:1998pc}
X.-D. Ji, \emph{{Off forward parton distributions}},
  \href{http://dx.doi.org/10.1088/0954-3899/24/7/002}{\emph{J. Phys. G} {\bf
  24} (1998) 1181--1205}, [\href{http://arxiv.org/abs/hep-ph/9807358}{{\tt
  hep-ph/9807358}}].

\bibitem{Diehl:2003ny}
M.~Diehl, \emph{{Generalized parton distributions}},
  \href{http://dx.doi.org/10.1016/j.physrep.2003.08.002}{\emph{Phys. Rept.}
  {\bf 388} (2003) 41--277}, [\href{http://arxiv.org/abs/hep-ph/0307382}{{\tt
  hep-ph/0307382}}].

\bibitem{Becher:2007cu}
T.~Becher and K.~Melnikov, \emph{{Two-loop QED corrections to Bhabha
  scattering}},
  \href{http://dx.doi.org/10.1088/1126-6708/2007/06/084}{\emph{JHEP} {\bf 06}
  (2007) 084}, [\href{http://arxiv.org/abs/0704.3582}{{\tt 0704.3582}}].

\bibitem{Becher:2010tm}
T.~Becher and M.~Neubert, \emph{{Drell-Yan Production at Small $q_T$,
  Transverse Parton Distributions and the Collinear Anomaly}},
  \href{http://dx.doi.org/10.1140/epjc/s10052-011-1665-7}{\emph{Eur. Phys. J.
  C} {\bf 71} (2011) 1665}, [\href{http://arxiv.org/abs/1007.4005}{{\tt
  1007.4005}}].

\bibitem{Becher:2011pf}
T.~Becher, G.~Bell and M.~Neubert, \emph{{Factorization and Resummation for Jet
  Broadening}},
  \href{http://dx.doi.org/10.1016/j.physletb.2011.09.005}{\emph{Phys. Lett. B}
  {\bf 704} (2011) 276--283}, [\href{http://arxiv.org/abs/1104.4108}{{\tt
  1104.4108}}].

\bibitem{Liu:2018czl}
T.~Liu and A.~Penin, \emph{{High-Energy Limit of Mass-Suppressed Amplitudes in
  Gauge Theories}},
  \href{http://dx.doi.org/10.1007/JHEP11(2018)158}{\emph{JHEP} {\bf 11} (2018)
  158}, [\href{http://arxiv.org/abs/1809.04950}{{\tt 1809.04950}}].

\bibitem{Kotsky:1997rq}
M.~I. Kotsky and O.~I. Yakovlev, \emph{{On the resummation of double logarithms
  in the process Higgs $\to\gamma\gamma$}},
  \href{http://dx.doi.org/10.1016/S0370-2693(97)01260-4}{\emph{Phys. Lett. B}
  {\bf 418} (1998) 335--344}, [\href{http://arxiv.org/abs/hep-ph/9708485}{{\tt
  hep-ph/9708485}}].

\bibitem{Bell:2008er}
G.~Bell and T.~Feldmann, \emph{{Modelling light-cone distribution amplitudes
  from non-relativistic bound states}},
  \href{http://dx.doi.org/10.1088/1126-6708/2008/04/061}{\emph{JHEP} {\bf 04}
  (2008) 061}, [\href{http://arxiv.org/abs/0802.2221}{{\tt 0802.2221}}].

\bibitem{Beneke:2019slt}
M.~Beneke, C.~Bobeth and R.~Szafron, \emph{{Power-enhanced leading-logarithmic
  QED corrections to $B_q \to \mu^+\mu^-$}},
  \href{http://dx.doi.org/10.1007/JHEP10(2019)232}{\emph{JHEP} {\bf 10} (2019)
  232}, [\href{http://arxiv.org/abs/1908.07011}{{\tt 1908.07011}}].

\bibitem{Beneke:2020vnb}
M.~Beneke, P.~B\"oer, J.-N. Toelstede and K.~K. Vos, \emph{{QED factorization
  of non-leptonic $B$ decays}},
  \href{http://dx.doi.org/10.1007/JHEP11(2020)081}{\emph{JHEP} {\bf 11} (2020)
  081}, [\href{http://arxiv.org/abs/2008.10615}{{\tt 2008.10615}}].

\end{thebibliography}\endgroup
\end{document}